\documentclass[a4paper,11pt,preprintnumbers]{article}
\pdfoutput=1

\usepackage{amssymb,mathrsfs,amsmath}
\usepackage{a4wide}
\usepackage{color,xcolor}
\usepackage{slashed,soul}
\usepackage{graphicx}
\usepackage{amsfonts}
\usepackage{lscape}
\def\linkcolor{cyan!70!black}

\usepackage[
colorlinks=true
,urlcolor=\linkcolor
,anchorcolor=\linkcolor
,citecolor=\linkcolor
,filecolor=\linkcolor
,linkcolor=\linkcolor
,menucolor=\linkcolor
,linktocpage=true
,pdfproducer=medialab
,pdfa=true
]{hyperref}
\usepackage{amsthm}
\usepackage{booktabs}
\usepackage{array}
\usepackage{rotating}
\usepackage[numbers, sort&compress]{natbib}
\usepackage{multirow}
\usepackage{float}
\usepackage[utf8]{inputenc}
\usepackage[T1]{fontenc}
\usepackage{appendix}
\usepackage{epsfig}
\usepackage{orcidlink}
\usepackage{comment}

\newcommand{\beq}{\begin{equation}} 
\newcommand{\eeq}{\end{equation}} 
\newcommand{\ba}{\begin{array}}  
\newcommand{\ea}{\end{array}} 
\newcommand{\bea}{\begin{eqnarray}}  
\newcommand{\eea}{\end{eqnarray} }  
\newcommand{\bal}{\begin{align}}
\newcommand{\eal}{\end{align}}   
\newcommand{\bi}{\begin{itemize}}  
\newcommand{\ei}{\end{itemize}}  
\newcommand{\ben}{\begin{enumerate}}  
\newcommand{\een}{\end{enumerate}}  
\newcommand{\bc}{\begin{center}}
\newcommand{\ec}{\end{center}} 
\newcommand{\bt}{\begin{table}}
\newcommand{\et}{\end{table}}  
\newcommand{\btb}{\begin{tabular}}
\newcommand{\etb}{\end{tabular}}

\newcommand{\Tr}[1]{\text{Tr}\left[#1\right]}
\newcommand{\abs}[1]{\left\lvert#1\right\rvert}

\newcommand{\SM}{\text{SM}}
\newcommand{\etaee}{\eta_{ee}}
\newcommand{\etamumu}{\eta_{\mu\mu}}
\newcommand{\etatautau}{\eta_{\tau\tau}}
\newcommand{\etaemu}{\eta_{e\mu}}
\newcommand{\etaetau}{\eta_{e\tau}}
\newcommand{\etamutau}{\eta_{\mu\tau}}
\newcommand{\Vud}{\abs{V_{ud}}}
\newcommand{\Vus}{\abs{V_{us}}}
\newcommand{\thetae}{\abs{\theta_e}}
\newcommand{\thetamu}{\abs{\theta_\mu}}
\newcommand{\thetatau}{\abs{\theta_\tau}}
\newcommand{\mlightest}{m_{\text{lightest}}}

\newcommand{\LFUratio}[3]{R_{#1 #2}^{#3}}

\def\arrvline{\hfil\kern\arraycolsep\vline\kern-\arraycolsep\hfilneg}
\newcommand{\sw}{s_{\mathrm{w}}}

\allowdisplaybreaks

\let\OLDthebibliography\thebibliography
\renewcommand\thebibliography[1]{
  \OLDthebibliography{#1}
  \setlength{\parskip}{0pt}
  \setlength{\itemsep}{0pt plus 0.3ex}
}

\begin{document}

\vspace{1cm}

\begin{titlepage}

\begin{flushright}
IFT-UAM/CSIC-23-60\\
FTUV-23-0531.7594\\
IFIC/23-19
 \end{flushright}
\vspace{0.2truecm}

\begin{center}
\renewcommand{\baselinestretch}{1.8}\normalsize
\boldmath
{\LARGE\textbf{
Bounds on lepton non-unitarity and heavy neutrino mixing
}}
\unboldmath
\end{center}

\vspace{0.4truecm}

\renewcommand*{\thefootnote}{\fnsymbol{footnote}}

\begin{center}

{
Mattias Blennow$^1$\footnote{\href{mailto:emb@kth.se}{emb@kth.se}}\orcidlink{0000-0001-5948-9152},
Enrique Fern\'andez-Mart\'inez$^2$\footnote{\href{mailto:enrique.fernandez-martinez@uam.es}{enrique.fernandez-martinez@uam.es}}\orcidlink{0000-0002-6274-4473}, 
Josu Hern\'andez-Garc\'ia$^3$\footnote{\href{mailto:garcia.josu.hernandez@ttk.elte.hu}{garcia.josu.hernandez@ttk.elte.hu}}\orcidlink{0000-0003-0734-0879},\\[1ex]
Jacobo L\'opez-Pav\'on$^4$\footnote{\href{mailto:Jacobo.Lopez@uv.es}{jacobo.lopez@uv.es}}\orcidlink{0000-0002-9554-5075},
Xabier Marcano$^2$\footnote{\href{mailto:xabier.marcano@uam.es}{xabier.marcano@uam.es}}\orcidlink{0000-0003-0033-0504}
and Daniel Naredo-Tuero$^2$\footnote{\href{mailto:daniel.naredo@uam.es}{daniel.naredo@uam.es}}\orcidlink{0000-0002-5161-5895}
}

\vspace{0.7truecm}

{\footnotesize
$^1$Department of Physics, School of Engineering Sciences, KTH Royal Institute of Technology, \\
AlbaNova University Center, Roslagstullsbacken 21, SE-106 91 Stockholm, Sweden
\\[.5ex]
$^2$Departamento de F\'{\i}sica Te\'orica and Instituto de F\'{\i}sica Te\'orica UAM/CSIC,\\
Universidad Aut\'onoma de Madrid, Cantoblanco, 28049 Madrid, Spain
\\[.5ex]
$^3$Institute for Theoretical Physics, ELTE E\"otv\"os Lor\'and University,\\ P\'azm\'any P\'eter s\'et\'any 1/A, H-1117 Budapest, Hungary
\\[.5ex]
$^4$ Instituto de F\'{\i}sica Corpuscular, Universidad de Valencia and CSIC\\ 
 Edificio Institutos Investigaci\'on, Catedr\'atico Jos\'e Beltr\'an 2, 46980 Spain

}

\vspace*{2mm}
\end{center}

\renewcommand*{\thefootnote}{\arabic{footnote}}
\setcounter{footnote}{0}

\begin{abstract}

We present an updated and improved global fit analysis of current flavor and electroweak precision observables to derive bounds on unitarity deviations of the leptonic mixing matrix and on the mixing of heavy neutrinos with the active flavours. 
This new analysis is motivated by new and updated experimental results on key observables such as $V_{ud}$, the invisible decay width of the $Z$ boson and the $W$ boson mass. It also improves upon previous studies by considering the full correlations among the different observables and explicitly calibrating the test statistic, which may present significant deviations from a $\chi^2$ distribution. 
The results are provided for three different Type-I seesaw scenarios: the minimal scenario with only two additional right-handed neutrinos, the next to minimal one with three extra neutrinos, and the most general one with an arbitrary number of heavy neutrinos that we parametrize via a generic deviation from a unitary leptonic mixing matrix. Additionally, we also analyze the case of generic deviations from unitarity of the leptonic mixing matrix, not necessarily induced by the presence of additional neutrinos. This last case relaxes some correlations among the parameters and is able to provide a better fit to the data. Nevertheless, inducing only leptonic unitarity deviations avoiding both the correlations implied by the right-handed neutrino extension as well as more  strongly constrained operators is challenging and would imply significantly more complex UV completions.

\end{abstract}

\end{titlepage}

\tableofcontents

\section{Introduction}
Adding right-handed (RH) neutrinos to the Standard Model (SM) particle spectrum is one of the best motivated extensions to address several of the open problems in particle physics.
They constitute the simplest possibility to accommodate the evidence for neutrino masses and mixing, and they can also play an important role as dark matter candidates~\cite{Dodelson:1993je,Shi:1998km,Abazajian:2001nj,Asaka:2005an} or portals to dark sectors~\cite{Falkowski:2009yz,Lindner:2010rr,GonzalezMacias:2015rxl,Blennow:2019fhy} as well as in the generation of the matter-antimatter asymmetry of the Universe~\cite{Fukugita:1986hr,Akhmedov:1998qx,Asaka:2005pn,Drewes:2017zyw}. 

Due to the singlet nature of the RH neutrinos, also known as heavy neutral leptons (HNLs), there is a new energy scale associated to their allowed Majorana mass, which is unknown from the theoretical point of view. 
If this scale is low enough, they could be produced at different laboratories, motivating the strong experimental effort (see for instance~\cite{Antel:2023hkf}) that led to a plethora of constraints (see {\it e.g.}~\cite{Fernandez-Martinez:2023phj,MatheusRepository}).
On the other hand, if the scale is above the experimental energy, the RH neutrinos cannot be produced, but their existence induces deviations from unitarity of the leptonic mixing matrix, so they can still be probed for at the intensity frontier via electroweak precision observables, universality ratios or charged lepton flavor violating (cLFV) processes~\cite{Petcov:1976ff,Bilenky:1977du,Cheng:1977vn,Marciano:1977wx,Lee:1977qz,Lee:1977tib,Shrock:1980vy,Schechter:1980gr,Shrock:1980ct,Shrock:1981wq,Langacker:1988ur,Pilaftsis:1992st,Ilakovac:1994kj,Nardi:1994iv,Tommasini:1995ii,Illana:2000ic,Loinaz:2003gc,Arganda:2004bz,Loinaz:2004qc,Antusch:2006vwa,Antusch:2008tz,Biggio:2008in,Alonso:2012ji,Abada:2012mc,Akhmedov:2013hec,Basso:2013jka,Abada:2013aba,Arganda:2014dta,Antusch:2014woa,Antusch:2015mia,Abada:2014cca,Abada:2015oba,Abada:2015trh,Abada:2016awd,DeRomeri:2016gum,Arganda:2016zvc,Herrero:2018luu,Marcano:2019rmk,Chrzaszcz:2019inj,Coutinho:2019aiy,Calderon:2022alb}.

When the RH neutrino masses are above the experimental energy, it is useful to consider an effective field theory (EFT) parametrization of their impact at accessible energies.
After integrating them out, the lowest order dim-5 Weinberg operator~\cite{Weinberg:1979sa} generates neutrino masses and mixings after electroweak symmetry breaking, while the only dim-6 induced at tree level leads to non-unitarity effects in lepton mixing~\cite{Broncano:2002rw}.
In a high-scale seesaw mechanism~\cite{Minkowski:1977sc,Mohapatra:1979ia,Yanagida:1979as,Gell-Mann:1979vob}, the naive expectation is that both dim-5 and 6 operators are suppressed, explaining the smallness of neutrino masses at the price of rendering the model virtually untestable. 
On the other hand, symmetry protected~\cite{Branco:1988ex,Kersten:2007vk,Abada:2007ux,Moffat:2017feq} low-scale seesaw mechanisms, such as the linear~\cite{Mohapatra:1986aw,Mohapatra:1986bd} or inverse~\cite{Akhmedov:1995ip,Malinsky:2005bi} seesaw mechanisms, suppress the dim-5 operator, but not the dim-6 one, allowing for large deviations from unitarity. 

When extending the SM with $n$ new RH neutrinos $N_R^i$, both a neutrino Yukawa and a Majorana mass term for the right-handed neutrinos should be added to the Lagrangian: 
\begin{equation}
    \mathcal L =  
    - \frac12 \overline{N_R^{i\,c}} M_M^{ij} N_R^j
    - Y_\nu^{\alpha i} \overline{L_\alpha} \tilde{\phi} N^{i}_R
    + h.c.
\end{equation}
where $\tilde\phi = i\sigma_2\phi$ stands for the SM Higgs doublet, $Y_\nu$ is a complex $3\times n$ Yukawa coupling matrix and $M_M$ is a $n\times n$ symmetric Majorana matrix for the $N_R$ fields.
When the RH neutrinos are heavy, upon integrating them out, they induce light neutrino masses via the dim-5 Weinberg operator~\cite{Weinberg:1979sa},
\begin{equation}
    m_{\rm \nu} = -\Theta M_M \Theta^T\,,
    \label{eq:d=5}
\end{equation}
where $\Theta$ is the mixing between the active SM neutrinos and the RH ones:
\begin{equation}
    \Theta\equiv \dfrac{v}{\sqrt{2}}Y_\nu M_M^{-1} = m_D M_M^{-1}\,,
\end{equation}
and $v=246$ GeV is the Higgs {\it vev}. 
The symmetric mass matrix for the light neutrinos $m_\nu$ will be diagonalized by a unitary matrix $U$. Additionally, as already mentioned, the only dimension 6 operator generated at tree level induces the non-unitarity effects we are interested in~\cite{Broncano:2002rw}. In particular, the leptonic mixing matrix coupling the light neutrino mass eigenstates to the SM charged leptons through CC interactions will be given by:
\begin{equation}\label{eq:eta}
N = (\mathbb{I}-\eta)\, U\,.
\end{equation}
We have dubbed the leptonic mixing matrix $N$ to emphasize its \emph{non-unitary} character. Indeed, while $U$ is the unitary matrix that diagonalizes the Weinberg $d=5$ operator in Eq.~\eqref{eq:d=5}, $\eta$ parametrizes the unitarity deviations of $N$ and corresponds to (half of) the coefficient of the $d=6$ operator obtained upon integrating out the heavy fields~\cite{Broncano:2002rw}:
\begin{equation}\label{eq:eta_theta}
    \eta = \dfrac{1}{2}\Theta\Theta^\dagger\,.
\end{equation}

Notice that, any general matrix can be parametrized through the product of an Hermitian and a unitary matrix. Hence, Eq.~\eqref{eq:eta} is a completely general and convenient way to encode unitarity deviations through the small Hermitian matrix $\eta$~\cite{Fernandez-Martinez:2007iaa}. As we will show in more detail later, most observables depend on the combination $NN^\dagger$, since the flavor indices corresponding to the charged leptons are fixed by the process and the physical neutrino indices are summed over, since they are not measured. 
Being $\eta$ a small parameter, we have
\begin{equation}
    \sum_{i=1}^3 N^{\phantom\dagger}_{\alpha i}N^\dagger_{i\beta }=\delta_{\alpha\beta}-2\eta_{\alpha\beta}+\mathcal O\big(\eta_{\alpha\beta}^2\big)\,.
\end{equation}
Thus, the observables do not depend on $U$ and the $\eta$ matrix encodes any possible effect arising from non-unitarity, regardless of the UV completion that originates the deviations and how many fields it contains. 

In all generality, the dim-5 and 6 operators are independent, meaning that in the most general case we can decouple $\eta$ from light neutrino masses and mixing. Nevertheless,  when induced by RH neutrinos, $\eta$ is a positive-definite matrix, as shown by Eq.~\eqref{eq:eta_theta}, and hence its parameters are not entirely independent. In particular, the Schwarz inequality will be satisfied:  
\begin{equation}
    \abs{\eta_{\alpha\beta}}\leq \sqrt{\eta_{\alpha\alpha}\eta_{\beta\beta}}\,.
    \label{eq:schwarz}
\end{equation}
Furthermore, since both operators are generated from different combinations of the same Yukawa and mass matrices (see Eqs.~\eqref{eq:d=5} and~\eqref{eq:eta_theta}), it is possible to find correlations between them depending on how many new heavy neutrinos are considered. Thus, with the aim of covering from the minimal to the most general case, we will consider the following 4 scenarios:
\begin{itemize}
\item The minimal scenario that accommodates oscillation data adding only 2 RH neutrinos. Here the dim-5 and 6 operators are fully correlated, since the latter (and thus $\eta$) can be fully reconstructed from the former up to a global scale~\cite{Gavela:2009cd}.

\item The next to minimal scenario with 3 RH neutrinos, where the dim-5 operator still imposes strong correlations on $\eta$~\cite{Fernandez-Martinez:2015hxa}.

\item The most general RH neutrino scenario where we assume a general $\eta$ matrix, independent of neutrino oscillation data, which encodes the low-energy effects of an arbitrary number of heavy RH neutrinos. 

\item Generic deviations from unitarity not requiring $\eta$ to be positive-definite or subject to the Schwarz inequality. This would imply more elaborate additions to the SM particle content beyond only RH neutrinos (see Ref.~\cite{Coutinho:2019aiy} for a dedicated discussion).

\end{itemize}

In order to obtain a complete picture of the experimental situation, we will perform a global fit analysis of an extended set of observables that are affected by a non-unitary leptonic mixing matrix, and derive current limits for all these scenarios.
Compared to previous analyses~\cite{Antusch:2014woa,Fernandez-Martinez:2016lgt,Chrzaszcz:2019inj} we improve and extend the study in several ways:

\begin{itemize}

\item We update and complete the list of observables taking into account all correlations among them consistently. We also comment on the role of the CDF-II anomalous measurement of the $W$ mass~\cite{CDF:2022hxs}, given the importance of this observable in the non-unitarity analysis.

\item We extend the number of scenarios considered. In particular and as mentioned above, we specifically consider the minimal case with two heavy neutrinos and obtain its corresponding bounds, which is an analysis that was previously missing in the literature.  

\item We improve the statistical treatment with an explicit calibration of the test statistic, since deviations from Wilks' theorem are expected for most of the scenarios under consideration.

\item We avoid using the values provided in the PDG review for the SM prediction of the different observables. Indeed, while this procedure is common and convenient, these values are obtained from a global fit  which includes also the observables of interest in this study. This approach is therefore not consistent as it double counts these observables and, by using them in the prediction, it may artificially reduce any preference for new physics. We instead choose an independent and accurate set of input observables distinct from the others to make our predictions consistently.  

\end{itemize}

This paper is organized as follows. In section~\ref{Sec:observables} we describe the observables considered in our global fits specifying their dependence on the unitarity deviations of the leptonic mixing matrix and how we include them in our statistical analysis. In sections~\ref{Sec:Results2N}, \ref{Sec:Results3N} and~\ref{Sec:ResultsGSS} we present the results for the scenarios with 2, 3 and a general number of additional RH neutrinos respectively, while in section~\ref{Sec:ResultsGNU} we show the impact of dropping the assumption of $\eta$ being positive definite as it might be the case if not induced (only) by mixing with RH neutrinos. Finally, we summarize our conclusions in section~\ref{Sec:Discussion}.

%
\section{Observables}
\label{Sec:observables}

In this work, we will derive global bounds on deviations from unitarity, and consequently on heavy neutrino mixing, via a fit to the following observables:

\begin{itemize}
    \item[$\bullet$] Four determinations of the W-boson mass: $M_W^{\text{LEP}}$, $M_W^{\text{Tev}}$, $M_W^{\text{LHCb}}$, $M_W^{\text{ATLAS}}$.
    \item[$\bullet$] Two determinations of the effective weak angle: $s^{2\hspace{0.15cm}\text{LHC}}_{\text{eff}}$ and $s^{2\hspace{0.15cm}\text{Tev}}_{\text{eff}}$.
    \item[$\bullet$] Five LEP observables measured at the Z-pole, plus a determination of the $Z$ invisible width from CMS: $\Gamma_Z$, $\sigma^{0}_{\text{had}}$, $R_e$, $R_\mu$, $R_\tau$, $\Gamma_{\text{inv}}^{\text{LHC}}$.
    \item[$\bullet$] Five weak decay ratios constraining lepton flavor universality: $\LFUratio{\mu}{e}{\pi}$, $\LFUratio{\tau}{\mu}{\pi}$, $\LFUratio{\mu}{e}{K}$, $\LFUratio{\mu}{e}{\tau}$, $\LFUratio{\tau}{\mu}{\tau}$.
    \item[$\bullet$] Ten weak decays constraining CKM unitarity.
    \item[$\bullet$] cLFV observables.
\end{itemize}

With the obvious exception of cLFV processes, all these observables are lepton flavor conserving (LFC) and thus constrain the diagonal entries of the $\eta$ matrix. Nevertheless, through the Schwarz inequality of Eq.~\eqref{eq:schwarz}, LFC observables will also impose constraints on the off-diagonal entries. 
As we will see, these limits will be often stronger than those imposed by direct contributions to cLFV processes. 

\subsection*{SM input parameters and predictions}
For the SM input parameters the very precise determinations of the Z-boson mass $M_Z$, the fine structure constant $\alpha$ and the Fermi constant ($G_F$) extracted from $\mu$ decay $G_\mu$ will be used. The corresponding values are taken from the PDG \cite{ParticleDataGroup:2022pth}:
\begin{equation}
    \begin{aligned}
        M_Z&=91.1876(21)\text{ GeV}\,,
        \\
        \alpha^{-1}&=137.035999180(10)\,,\\
        G_\mu&=1.1663788(6)\cdot 10^{-5} \text{ GeV$^{-2}$}\,.
    \end{aligned}
\end{equation}
However, since the extraction of $G_F$ comes from a charged current weak decay and the $W$ boson couplings to leptons are modified due to the non-unitarity of the leptonic mixing matrix, $G_\mu$ will no longer be equal to $G_F$. In fact, in the presence of a non-unitary mixing matrix the muon decay width gets modified as:
\begin{equation}
    \Gamma_\mu=\frac{G_F^2m_\mu^5}{192\pi^3}\sum_{i=1}^3\left\lvert N_{\mu i}\right\rvert^2 \sum_{j=1}^3\left\lvert N_{ej}\right\rvert^2 \simeq \frac{G_F^2m_\mu^5}{192\pi^3}\left(1-2\eta_{ee}-2\eta_{\mu\mu}\right)\equiv \frac{G_\mu^2m_\mu^5}{192\pi^3}\,,
\end{equation}
and thus $G_F$, as measured through muon decay, picks a non-unitarity dependence, 
\begin{equation}
    G_F\simeq G_\mu\left(1+\etaee+\etamumu\right),
\end{equation}
which will propagate into any observable which depends on it.
In the following, we discuss in detail the dependence of these observables on the non-unitarity parameters. The main results are already summarised in Table~\ref{table:observables}.

The SM predictions for all the observables entering in our fit will also be necessary. 
A common practice is to take their SM values as given by the PDG, however this would not be consistent since the PDG values correspond to a global fit including the electroweak observables themselves. Hence, since our goal is precisely to perform such a fit but in the context of non-unitarity, it is not consistent to take values which have already been fitted. Moreover, since the experimental data is being fitted, the resulting SM predictions would be systematically closer to the experimental values than the values obtained by just inserting the SM input parameters into the expression for the observables. This may artificially diminish any preference for new physics in the data.

For this reason, we will instead derive the SM predictions for each observable from its parametric dependence on the SM input parameters.
This implies including the relevant loop corrections, which requires additional inputs: the mass of the top quark ($m_t$), the mass of the Higgs boson ($M_H$), the strong coupling constant at $M_Z$ ($\alpha_s\left(M_Z\right)$) and the running of the fine-structure constant at $M_Z$ ($\Delta\alpha\left(M_Z\right)$). We adopt the following values~\cite{ParticleDataGroup:2022pth}:
\begin{equation}
    \begin{aligned}
        m_t&=172.69(30) \text{ GeV}\,,\\
        M_H&=125.25(17) \text{ GeV}\,,\\
        \alpha_s\left(M_Z\right)&=0.1185(16)\,,\\
        \Delta\alpha\left(M_Z\right)&=591.05(70)\cdot10^{-5}\,.
    \end{aligned}
\end{equation}
Notice that we only consider SM loop contributions, which are of course necessary for the precise evaluation of the SM prediction of each observable. 
In principle one should also take into account heavy neutrino loop contributions to the non-unitarity dependence of the observables, however it has been shown that these can be safely neglected as they are subleading~\cite{Fernandez-Martinez:2015hxa}.

\begin{table}[h!]
\small
\centering
    \begin{tabular}{|c|c|cc|}
         \hline &&&\\[-2ex]
        Observable & SM prediction & \multicolumn{2}{c|}{Experimental value}\\[0.5ex]
        \hline\hline
        &&&\\[-2ex]
        $M_W\simeq M_W^{\SM}\left(1+0.20\left(\eta_{ee}+\eta_{\mu\mu}\right)\right)$
        & $80.356(6)$ GeV & $80.373(11)$ GeV
        &-\\ [1.2ex]        
        $s^{2\hspace{0.15cm}\text{Tev}}_{\text{eff}}\simeq s^{2\hspace{0.15cm}\SM}_{\text{eff}}\left(1-1.40\left(\eta_{ee}+\eta_{\mu\mu}\right)\right)$
        &$0.23154(4)$&$0.23148(33)$
        &\cite{ParticleDataGroup:2022pth}\\[1.2ex]

        $s^{2\hspace{0.15cm}\text{LHC}}_{\text{eff}}\simeq s^{2\hspace{0.15cm}\SM}_{\text{eff}}\left(1-1.40\left(\eta_{ee}+\eta_{\mu\mu}\right)\right)$
        &$0.23154(4)$&$0.23129(33)$
        &\cite{ParticleDataGroup:2022pth}\\[1.2ex]
        \hline\hline
        &&&\\[-2ex]
        $\Gamma_{\text{inv}}^{\text{LHC}}\simeq\Gamma_{\text{inv}}^{\SM}\left(1-0.33\left(\eta_{ee}+\eta_{\mu\mu}\right)-1.33\eta_{\tau\tau}\right)$
        &$0.50145(5)$ GeV&$0.523(16)$ GeV
        &\cite{CMS:2022ett} \\ [1.2ex]
        $\Gamma_Z\simeq\Gamma_{Z}^{\SM}\left(1+1.08\left(\eta_{ee}+\eta_{\mu\mu}\right)-0.27\eta_{\tau\tau}\right)$
        &$2.4939(9)$ GeV&$2.4955(23)$ GeV
        &\cite{ParticleDataGroup:2022pth} \\ [1.2ex]
        $\sigma^{0}_{\text{had}}\simeq\sigma^{0\hspace{0.15cm}\SM}_{\text{had}}\left(1+0.50\left(\eta_{ee}+\eta_{\mu\mu}\right)+0.53\eta_{\tau\tau}\right)$
        &$41.485(8)$ nb&$41.481(33)$ nb
        &\cite{ParticleDataGroup:2022pth} \\ [1.2ex]
        
        $R_e\simeq R_e^{\SM}\left(1+0.27\left(\eta_{ee}+\eta_{\mu\mu}\right)\right)$
        &$20.733(10)$&$20.804(50)$
        &\cite{ParticleDataGroup:2022pth}\\[1.2ex]
        
        $R_\mu\simeq R_\mu^{\SM}\left(1+0.27\left(\eta_{ee}+\eta_{\mu\mu}\right)\right)$
        &$20.733(10)$&$20.784(34)$
        &\cite{ParticleDataGroup:2022pth}\\[1.2ex]
        
        $R_\tau\simeq R_\tau^{\SM}\left(1+0.27\left(\eta_{ee}+\eta_{\mu\mu}\right)\right)$&$20.780(10)$&$20.764(45)$
        &\cite{ParticleDataGroup:2022pth}\\[1.2ex]
        
        \hline\hline
        &&&\\[-2ex]
        
        $\LFUratio{\mu}{e}{\pi}\simeq \left(1-\left(\eta_{\mu\mu}-\eta_{ee}\right)\right)$
        &$1$&$1.0010(9)$
        &\cite{Bryman:2021teu}\\[1.2ex]
        
        $\LFUratio{\tau}{\mu}{\pi}\simeq \left(1-\left(\eta_{\tau\tau}-\eta_{\mu\mu}\right)\right)$
        &$1$&$0.9964(38)$
        &\cite{Bryman:2021teu}\\[1.2ex]
        
        $\LFUratio{\mu}{e}{K}\simeq \left(1-\left(\eta_{\mu\mu}-\eta_{ee}\right)\right)$
        &$1$&$0.9978(18)$
        &\cite{Bryman:2021teu}\\[1.2ex]
        
        $\LFUratio{\mu}{e}{\tau}\simeq \left(1-\left(\eta_{\mu\mu}-\eta_{ee}\right)\right)$
        &$1$&$1.0018(14)$
        &\cite{Bryman:2021teu}\\[1.2ex]
        
        $\LFUratio{\tau}{\mu}{\tau}\simeq \left(1-\left(\eta_{\tau\tau}-\eta_{\mu\mu}\right)\right)$
        &$1$&$1.0010(14)$
        &\cite{Bryman:2021teu}\\[1.2ex]
        
        \hline\hline 
        &&&\\ [-2ex]
        $\abs{V_{ud}^\beta}\simeq \sqrt{1-\abs{V_{us}}^2}\left(1+\eta_{\mu\mu}\right)$
        &$\sqrt{1-\abs{V_{us}}^2}$&$0.97373(31)$
        &\cite{ParticleDataGroup:2022pth}\\[1.2ex]
        
        $\abs{V_{us}^{\tau\rightarrow K\nu}}\simeq \abs{V_{us}}\left(1+\eta_{ee}+\eta_{\mu\mu}-\eta_{\tau\tau}\right)$&$\abs{V_{us}}$
        &$0.2236(15)$&\cite{HFLAV:2019otj}\\[1.2ex]
        
        $\abs{V_{us}^{\tau\rightarrow K,\pi}}\simeq \abs{V_{us}}\left(1+\eta_{\mu\mu}\right)$
        &$\abs{V_{us}}$&$0.2234(15)$
        &\cite{ParticleDataGroup:2022pth}\\[1.2ex]
        
        $\abs{V_{us}^{K_L\rightarrow \pi e \nu}}\simeq \abs{V_{us}}\left(1+\eta_{\mu\mu}\right)$
        &$\abs{V_{us}}$&$0.2229(6)$
        &\cite{ParticleDataGroup:2022pth}\\[1.2ex]
        
        $\abs{V_{us}^{K_L\rightarrow \pi \mu \nu}}\simeq \abs{V_{us}}\left(1+\eta_{ee}\right)$
        &$\abs{V_{us}}$&$0.2234(7)$
        &\cite{ParticleDataGroup:2022pth}\\[1.2ex]
        
        $\abs{V_{us}^{K_S\rightarrow \pi e \nu}}\simeq \abs{V_{us}}\left(1+\eta_{\mu\mu}\right)$
        &$\abs{V_{us}}$&$0.2220(13)$
        &\cite{ParticleDataGroup:2022pth}\\[1.2ex]
        
        $\abs{V_{us}^{K_S\rightarrow \pi\mu\nu}}\simeq \abs{V_{us}}\left(1+\eta_{ee}\right)$
        &$\abs{V_{us}}$&$0.2193(48)$
        &\cite{ParticleDataGroup:2022pth}\\[1.2ex]
        
        $\abs{V_{us}^{K^{\pm}\rightarrow \pi e \nu}}\simeq \abs{V_{us}}\left(1+\eta_{\mu\mu}\right)$
        &$\abs{V_{us}}$&$0.2239(10)$
        &\cite{ParticleDataGroup:2022pth}\\[1.2ex]
        
        $\abs{V_{us}^{K^{\pm}\rightarrow \pi \mu \nu}}\simeq \abs{V_{us}}\left(1+\eta_{ee}\right)$
        &$\abs{V_{us}}$&$0.2238(12)$
        &\cite{ParticleDataGroup:2022pth}\\[1.2ex]
        
        $\abs{\dfrac{V_{us}}{V_{ud}}}^{K,\pi\rightarrow \mu \nu}\simeq \dfrac{\abs{V_{us}}}{\sqrt{1-\abs{V_{us}}^2}}$
        &$\dfrac{\abs{V_{us}}}{\sqrt{1-\abs{V_{us}}^2}}$
        &$0.23131(53)$
        &\cite{ParticleDataGroup:2022pth}\\[3.5ex]
        \hline
    \end{tabular}
    \caption{
    Set of precision observables used as input for our global fit. The first column includes their dependence on the non-unitarity parameters, the second their SM prediction and the third one their experimental value. 
    For $M_W$, we compute the current world average without including the CDF-II measurement (see text for details). }
    \label{table:observables}
\end{table}

\subsection[Constraints from $M_W$ and $s^{2}_{\text{eff}}$]{Constraints from $\boldsymbol{M_W}$ and $\boldsymbol{s^{2}_{\text{eff}}}$}

Although $M_W$ and $\sw^2$ do not depend directly on the non-unitarity parameters, they provide alternative determinations of $G_F$ to be compared to the one extracted from muon decay, which, as previously discussed, does depend on $\eta$.
In other words, the predicted values for $M_W$ and $\sw^2$ through $G_\mu$ will inherit the following dependence on $\eta$
\begin{align}
    M_W&=M_Z\sqrt{\dfrac{1}{2}+\sqrt{\dfrac{1}{4}-\dfrac{\pi\alpha\left(1+\etaee+\etamumu\right)}{\sqrt{2}G_\mu M_Z^2\left(1-\Delta r\right)}}}\,,\label{eq:MW}\\
    \sw^2&=\dfrac{1}{2}\left(1-\sqrt{1-\dfrac{2\sqrt{2}\pi \alpha\left(1+\etaee+\etamumu\right)}{G_\mu M_Z^2\left(1-\Delta r\right)}}\right)\,,\label{eq:sW}
\end{align}
with the SM radiative corrections included in $\Delta r=0.03657 (22)$~\cite{ParticleDataGroup:2022pth}.
Here, it is useful to factorize the SM contribution and keep only the leading corrections in $\eta$, obtaining the expressions reported in Table~\ref{table:observables}.
Then, we can compute more precisely the SM predictions for $M_W$ and $s_{\text{eff}}^2$ using, respectively, the formulas given in \cite{Awramik:2003rn} and \cite{Awramik:2006uz}.
Notice that the ``on-shell'' value of $\sw^2$ is related to the effective weak angle $s_{\text{eff}}^2$ in the table through a multiplicative factor $\kappa$ that includes SM loops~\cite{ParticleDataGroup:2022pth} 
\begin{equation}
    s_{\text{eff}}^2=\kappa\, \sw^2\,,\label{eq:seff}
\end{equation}
and, therefore, it has the same $\eta$-dependence than $\sw^2$.

These $\eta$-dependent predictions are to be compared with the current experimental values.  
For the $W$ boson mass, we compute the current world average also considering the most recent ATLAS result~\cite{ATLAS:2023fsi}. For this, we follow the PDG \cite{ParticleDataGroup:2022pth} prescription and combine the results of ATLAS, LHCb~\cite{LHCb:2021bjt}, Tevatron~\cite{CDF:2013dpa} and LEP~\cite{ALEPH:2013dgf} following the BLUE procedure~\cite{LYONS1988110,VALASSI2003391} and taking into account the corresponding correlations\footnote{Following the PDG~\cite{ParticleDataGroup:2022pth}, we will assume a correlated uncertainty of 9 MeV for LHCb and ATLAS in order to compute the LHC average. Then, assuming a correlation of 7 MeV between LHC and Tevatron, we compute the hadron collider average. The world average is then extracted by combining the hadron collider average with the LEP measurement assuming no correlations.}. In section~\ref{sec:DCF} we will discuss the latest and anomalous CDF-II measurement~\cite{CDF:2022hxs}. Since we find there is too much tension with other observables even in presence of non-unitarity, we will not include it in the global fit.

Regarding the weak mixing angle, we use two measurements of $s^2_{\text{eff}}$ coming from the LHC and Tevatron.
On the other hand, we do not include low-energy determinations of $\sw^2$, such as its determination from M\o ller scattering. This is due to the fact that in the electron vector coupling $g_V^{e}=-\frac{1}{2}+2\sw^2$ there is a partial cancellation (since $\sw^2\sim 1/4$). This renders the extraction of $\sw^2$ very sensitive to higher order contributions which would require a higher level of accuracy to include it reliably in the fit.

\subsection{Constraints from Z-pole observables}

All of the predicitions for the Z-pole observables will be modified, as they depend on $G_F$ and $\sw^2$. For instance, this is the case for the partial decay widths to charged fermions:
\begin{equation}
    \Gamma_{f}=N_C\dfrac{G_\mu M_Z^3\left(g_V^{f2}+g_A^{f2}\right)}{6\sqrt{2}\pi}\left(1+\etaee+\etamumu\right),
\end{equation}
where $N_C$ is a color factor and $g_V^f$ and $g_A^f$ are the vector and axial couplings of the $Z$ boson to fermions, respectively:
\begin{equation}
    \begin{aligned}
        g_V^f&=T_f-2Q_f\sw^2\,,\\
        g_A^f&=T_f\,,
    \end{aligned}
\end{equation}
with $T_f$ and $Q_f$ being the isospin and charge of the fermion. 

Moreover, the $Z$ boson invisible decay width will be also modified when the right-handed neutrinos are heavier than $M_Z$ and not kinematically available: 
\begin{equation} 
    \Gamma_{\text{inv}}=\dfrac{G_\mu M_Z^3\sum_{i,j=1}^3\Big|\big(N^\dagger N\big) _{ij}\Big|^2}{12\sqrt{2}\pi}\left(1+\etaee+\etamumu\right)\simeq\dfrac{G_\mu M_Z^3}{12\sqrt{2}\pi}\Big(3-\big(\etaee+\etamumu+4\etatautau\big)\Big)\,.
\end{equation}

Thus, non-unitarity effects reduce the $Z$ invisible decay width with respect to the SM prediction which, interestingly, helped explaining the long-standing $\sim2\sigma$ LEP anomaly in the number of neutrinos.
Nevertheless, this anomaly is now gone after an improvement in the computation of the Bhabha scattering cross section \cite{Janot:2019oyi}. On top of that, a recent determination of $\Gamma_{\text{inv}}$ from the CMS experiment~\cite{CMS:2022ett} is $\sim 1\sigma$ away from the SM towards larger values of $\Gamma_{\text{inv}}$, which imposes further constraints on non-unitarity effects.

Having this in mind, the observables of our fit will be the usual combinations used for LEP from the partial widths of the Z. Namely, we have the total width $\Gamma_Z$, the hadronic cross section $\sigma^{0}_{\text{had}}$ and the leptonic ratios $R_{\ell}$, defined as:
\begin{align}
    \Gamma_Z&=\sum_{f\neq t}\Gamma_f+\Gamma_{\text{inv}}\,,\\
\sigma^0_{\text{had}}&=\dfrac{12\pi\Gamma_e\Gamma_{\text{had}}}{M_Z^2\Gamma_Z^2}\,,\\
R_\ell &=\dfrac{\Gamma_\ell}{\Gamma_{\text{had}}}\,,
\end{align}
where $\Gamma_{\text{had}}=\sum_{q\neq t} \Gamma_{q}$. 
Moreover, we will include the new CMS measurement of $\Gamma_{\rm inv}$ as an additional independent observable.

The parametric dependences of the different $Z$ partial widths in order to extract the SM predictions are taken from \cite{Freitas:2014hra}. The expression for the singlet vector contribution to the full hadronic width of the $Z$ is taken from \cite{Baikov:2012er}.
It should be noted that, since all of these observables (except $\Gamma_{\text{inv}}^{\text{LHC}}$) were measured in LEP, correlations between them must be taken into account, which we have taken from Table B.13 of~\cite{Janot:2019oyi}. 

Finally, it is worth mentioning that there exist other LEP observables, such as $R_c$, $R_b$ or the forward-backward asymmetries $A_{\rm FB}$, which are not included here. The reason why we have chosen to leave them out is the fact that they are not as precisely measured as other observables that depend on the same combination of $\eta$-parameters, thus they hold less constraining power and would end up diluting the goodness-of-fit by artificially increasing the number of degrees of freedom.

\subsection{Constraints from lepton flavor universality (LFU)}

The tightest constraints on the universality of weak interactions among the different lepton flavours come from ratios of meson or charged lepton decay widths differing in the flavor of one of the leptons involved. Since they are ratios, the $\eta$ dependence coming from $G_F$ will cancel out, and only the contributions coming from the $W$ vertex will matter. Namely:
\begin{align}
    \dfrac{\Gamma\left(P\rightarrow \mu \nu\right)}{\Gamma\left(P\rightarrow e \nu\right)}&=\dfrac{\Gamma\left(P\rightarrow \mu \nu_\mu\right)^{\SM}}{\Gamma\left(P\rightarrow e \nu_e\right)^{\SM}}\dfrac{\sum_{i=1}^3\abs{N_{\mu i}}^2}{\sum_{i=1}^3\abs{N_{e i}}^2} \simeq \dfrac{\Gamma\left(P\rightarrow \mu \nu_\mu\right)^{\SM}}{\Gamma\left(P\rightarrow e \nu_e\right)^{\SM}}\left(R_{\mu e}^P\right)^2,
    \\
    \dfrac{\Gamma\left(\tau\rightarrow P \nu\right)}{\Gamma\left(P\rightarrow \mu \nu\right)}&=\dfrac{\Gamma\left(\tau\rightarrow P \nu_\tau\right)^{\SM}}{\Gamma\left(P\rightarrow \mu \nu_\mu\right)^{\SM}}\dfrac{\sum_{i=1}^3\abs{N_{\tau i}}^2}{\sum_{i=1}^3\abs{N_{\mu i}}^2} \simeq \dfrac{\Gamma\left(\tau\rightarrow P\nu_\tau\right)^{\SM}}{\Gamma\left(P\rightarrow \mu \nu_\mu\right)^{\SM}}\left(R_{\tau \mu}^P\right)^2,
\end{align}
where $P=\pi,K$ and the ratio $R^P_{\alpha\beta}$ is defined as:
\begin{equation}
    R_{\alpha\beta}^P=1-\left(\eta_{\alpha\alpha}-\eta_{\beta\beta}\right).
    \label{obs:LFUratio}
\end{equation}

Similarly, there are also competitive bounds coming from ratios of fully leptonic decays. In particular, the $\mu - e$ sector can be constrained via:
\begin{equation}
    \dfrac{\Gamma\left(\tau\rightarrow \mu \nu \nu\right)}{\Gamma\left(\tau\rightarrow e \nu \nu\right)}=\dfrac{\Gamma\left(\tau\rightarrow \mu \nu_\mu \nu_\tau\right)^{\SM}}{\Gamma\left(\tau\rightarrow e \nu_e\nu_\tau\right)^{\SM}}\dfrac{\sum_{i=1}^3\abs{N_{\mu i}}^2}{\sum_{i=1}^3\abs{N_{e i}}^2} \simeq \dfrac{\Gamma\left(\tau\rightarrow \mu \nu_\mu \nu_\tau\right)^{\SM}}{\Gamma\left(\tau\rightarrow e \nu_e\nu_\tau\right)^{\SM}}\left(R_{\mu e}^\tau\right)^2,
\end{equation}
whereas the $\tau-\mu$ sector is constrained by:
\begin{equation}
    \dfrac{\Gamma\left(\tau\rightarrow e \nu \nu\right)}{\Gamma\left(\mu\rightarrow e \nu \nu \right)}=\dfrac{\Gamma\left(\tau\rightarrow e \nu_e \nu_\tau\right)^{\SM}}{\Gamma\left(\mu\rightarrow e\nu_e\nu_\mu\right)^{\SM}}\dfrac{\sum_{i=1}^3\abs{N_{\tau i}}^2}{\sum_{i=1}^3\abs{N_{\mu i}}^2} \simeq \dfrac{\Gamma\left(\tau\rightarrow e \nu_e \nu_\tau\right)^{\SM}}{\Gamma\left(\mu\rightarrow e\nu_e\nu_\mu\right)^{\SM}}\left(R_{\tau\mu}^\tau\right)^2.
\end{equation}
Here $R^\tau_{\alpha\beta}$ has the same $\eta$-dependence as in Eq.~\eqref{obs:LFUratio}. The correlations between the ratios extracted from $\tau$ decays are taken from \cite{Bryman:2021teu}. Notice that $R^K_{\tau \mu}$ has not been included in Table~\ref{table:observables} to avoid double counting since these decays will be included individually as independent measurements of $V_{us}$ as described below.

\subsection{Constraints from CKM unitarity}

The unitarity of the first row of the CKM is also constrained with significant accuracy. While violations of the unitarity of the leptonic mixing matrix leave the CKM matrix unchanged, the processes by which the values of its elements are extracted involve weak decays and will inherit a dependence on the $\eta$ parameters.

Since the CKM matrix is still unitary, the following CKM unitarity relation holds 
\begin{equation}
    \Vud^2+\Vus^2+\abs{V_{ub}}^2=1\,,
    \label{CKM:unitarity}
\end{equation}
where, given the uncertainty on the value of $\Vus$, the element $\abs{V_{ub}}=3.82(20)\cdot 10^{-3}$  can be safely neglected and thus one can substitute:
\begin{equation}
    \Vud=\sqrt{1-\Vus^2}\,.
\end{equation}
We will treat $\Vus$ as a nuisance parameter which will be marginalised over.

The most precise determination of $\Vud$ comes from superallowed $\beta$ decays. As such, its extraction will be modified by the leptonic $W$ vertex by ($1-2\etaee$) and additionally by $G_F^2$ with ($1+2\etaee+2\etamumu$), which in the end amounts to:
\begin{equation}
    \abs{V^\beta_{ud}}=\sqrt{1-\Vus^2}\left(1+\etamumu\right).
    \label{obs:Vudfromsuperallowedbeta}
\end{equation}

On the other hand, $\Vus$ can be determined from semileptonic and leptonic $K$ decays, as well as $\tau$ decays.
For the case of the $K$ semileptonic decays, similarly to the superallowed $\beta$ decays, the $\eta$ dependence will come from the $W$ vertex and the indirect dependence of $G_F$. In particular:
    \begin{align}
        \Big|V^{K\rightarrow\pi e\nu\;}_{us}\Big|&=\Vus \left(1+\etamumu\right),\\
        \Big|V^{K\rightarrow\pi \mu\nu}_{us}\Big|&=\Vus \left(1+\etaee\right).
    \end{align}

Moreover, one can also constrain the ratio $\Vus/\Vud$ by means of the ratio between $K$ and $\pi$ leptonic decay widths:
\begin{equation}
    \abs{\dfrac{V_{us}}{V_{ud}}}^{K,\pi \rightarrow\mu\nu}=\dfrac{\Vus}{\sqrt{1-\Vus^2}}\,.
    \label{obs:Vus/VudfromKpi}
\end{equation}
In this case, there is no dependence on the $\eta$ parameters, as they cancel out due to the fact that both final states have the same flavor. However, this ratio is still useful to constrain the nuisance parameter $V_{us}$.

Regarding $\tau$ decays, $\Vus$ can be extracted from its decay to a $K$:
\begin{equation}
    \abs{V^{\tau\rightarrow K\nu}_{us}}=\Vus\left(1+\etaee+\etamumu-\etatautau\right),
\end{equation}
where the $\eta$ dependence arises from the $W$ vertex with the $\tau$ and from $G_F$. Additionally, another determination can be made from the ratio of the $\tau$ decay widths to $K$ and $\pi$. This ratio, as in the case of Eq.~\eqref{obs:Vus/VudfromKpi}, does not depend on $\eta$. However, as one needs to multiply it by the determination of $\Vud$ to get $\Vus$, it inherits the dependence of Eq.~\eqref{obs:Vudfromsuperallowedbeta}.
\begin{equation}
    \abs{V^{\tau\rightarrow K,\pi}_{us}}=\Vus\left(1+\etamumu\right).
\end{equation}

It should be noted that after a reassessment of radiative corrections to the neutrino and superallowed $\beta$ decays, the estimated value of $V_{ud}$ decreased, leading to a 4-$5 \sigma$ tension with the assumption of CKM unitarity known as the Cabibbo anomaly~\cite{Seng:2018yzq,Czarnecki:2019mwq,Seng:2020wjq,Grossman:2019bzp}. More recently, the uncertainty associated to this measurement has also been revised and increased due to a more conservative estimate of the nuclear structure uncertainties~\cite{ParticleDataGroup:2022pth}. This has reduced the Cabibbo anomaly to a 2-$3 \sigma$ effect. Nevertheless, for a positive-definite $\eta$-matrix, this anomaly is only worsened for $\etamumu>0$, as shown by Eq.~\eqref{obs:Vudfromsuperallowedbeta} and therefore will push the global fit to very small values of this parameter.

\subsection[About the CDF-II $M_W$ measurement]{About the CDF-II $\boldsymbol{M_W}$ measurement}\label{sec:DCF}

Given its accuracy, one of the most important constraints on the unitarity of the leptonic mixing matrix and heavy neutrino mixing is the comparison of $M_W$ with its value obtained from $G_F$ from muon decay, and therefore affected by $\eta_{ee}+\eta_{\mu\mu}$ as discussed above, see Eq.~\eqref{eq:MW}. 

However, the most recent (and most precise) determination of $M_W$ by the CDF-II collaboration~\cite{CDF:2022hxs} is around 7$\sigma$ larger than the SM and around $3\sigma$ with respect to the prior world average. Intriguingly, this larger value of $M_W$, in tension with the SM, can in principle be explained through non-unitarity parameters for positive definite $\eta_{ee}+\eta_{\mu\mu}$, as shown in Table~\ref{table:observables}. This possibility was explored in Ref.~\cite{Blennow:2022yfm} considering a subset of the observables studied here including in particular the invisible width of the $Z$, LFU ratios and tests of CKM unitarity. 
The conclusion of that study was that the combined fit to the CDF-II $M_W$ measurement plus the other observables was in significant tension, mainly due to the Cabibbo anomaly which, at the time, was estimated to be $4$-$5\sigma$ prior to the revision of the uncertainty in the determination of $V_{ud}$. Indeed, the Cabibbo anomaly would prefer negative values for $\eta$, and it can thus only be worsened when the CDF-II $M_W$ anomaly is accommodated. 

The recent revision of the significance of the Cabibbo anomaly to the $2$-$3\sigma$ level invites to reconsider the explanation of the CDF-II anomaly through non-unitarity. Nevertheless, we find that this measurement is still in too much tension with, not only the other determinations of $M_W$, but also with other very precise measurements such as $s_{\text{eff}}^2$ and the Z-pole observables. 
In fact, even though the tension between the CDF-II measurement of $M_W$ and its SM prediction could be explained by non-unitarity, $M_W$, $s_{\text{eff}}^2$ and the Z-pole observables all depend on the same combination of $\eta$ parameters ({\it i.e.}~$\etaee+\etamumu$). Thus, any tension between these measurements cannot be improved by the presence of non-unitarity. 

We quantify this tension through the parameter goodness-of-fit (p-g.o.f.)~\cite{Maltoni:2003cu}, which is particularly suited to explore the situation in which two or more sets of observables are in tension. In practice it amounts to splitting the dataset into two (or more) subsets, $A_1$ and $A_2$, and computing:
\begin{equation}
    \overline{\chi}^2=\chi^2_{12}-\chi^2_1-\chi^2_2\,,
\end{equation}
where $\chi^2_{i}$ is the minimum of the $\chi^2$ considering the dataset $A_i$, and $\chi^2_{12}$ is the $\chi^2$ minimum considering both datasets. 
In the following, we take $A_1$ as the CDF-II $M_W$ measurement alone and consider three other options for the second set of data:
\begin{itemize}
    \item[$\bullet$] $A_2$ as the dataset formed by $s^{2\hspace{0.15cm}\text{LHC}}_{\text{eff}}$,  $s^{2\hspace{0.15cm}\text{Tev}}_{\text{eff}}$ and the world average $M_W$ without CDF-II.
    \item[$\bullet$] $A_2$ as the Z-pole observables.
    \item $A_2$ as the combination of the two previous sets of data.
\end{itemize}

The results of these three p-g.o.f.~are summarised in Table~\ref{table:pgof}. 
As expected, adding the measurements of $s_{\text{eff}}^2$ increases the tension already present between the new CDF-II result and other determinations of $M_W$, pushing it above the 4$\sigma$ level. 
Moreover, the tension of CDF-II with the Z-pole observables alone is also above $4\sigma$, indicating a high incompatibility due to the fact that the Z-pole observables are in good agreement with the SM expectation. 
Finally, the combination of the non-CDF-II $M_W$ measurements, effective weak angle and Z-pole observables results in a tension above $5\sigma$, clearly indicating that the CDF-II measurement cannot be reconciled with the aforementioned observables through unitarity deviations of the leptonic mixing matrix, since they all share the same dependence on these parameters and, therefore, we choose not to include it among the observables of our global fit.

\begin{table}[t!]
\centering
    \begin{tabular}{|c|c|c|c|}
         \hline&&&\\[-2ex]
         Set of observables & p-g.o.f. & tension & p-value\\[0.5ex]
        \hline\hline
        &&&\\[-2ex]
        CDF-II vs $M_W/s^2_{\text{eff}}$&$21.75/1$& 4.7$\sigma$ & $3.1\cdot 10^{-6}$\\[1.2ex]
        CDF-II vs Z-pole&$21.48/1$& 4.6$\sigma$ & $3.6\cdot 10^{-6}$\\[1.2ex]
        CDF-II vs $M_W/s^2_{\text{eff}}$ and Z-pole&$27.30/1$& 5.2$\sigma$ & $1.7\cdot 10^{-7}$\\[1.2ex]
        \hline
    \end{tabular}
    \caption{Tension between different sets of observables, quantified through the parameter goodness-of-fit~\cite{Maltoni:2003cu}. Note that the CDF-II measurement of $M_W$ is in tension not only with other determinations of $M_W$ and $s_{\text{eff}}$, but also with the $Z$-pole observables measured at LEP.}
    \label{table:pgof}
\end{table}

\subsection{Charged Lepton Flavor Violation}
\label{Sec:cLFV}

In the presence of heavy neutrinos, charged lepton flavor violating processes are no longer protected by the GIM mechanism~\cite{Glashow:1970gm}, as both non-unitarity of the leptonic mixing matrix and the scale separation between light and heavy neutrinos prevent such cancellation. 
Therefore, given the strong experimental constraints on cLFV transitions, they can be used to derive bounds on the off-diagonal elements of the $\eta$-matrix.

Currently, the most relevant cLFV processes for probing heavy neutrinos include~\cite{Ilakovac:1994kj,Alonso:2012ji} radiative decays, three body leptonic decays and $\mu-e$ conversion in heavy nuclei, whose present bounds are summarised in Table~\ref{table:LFV}.
The available list of cLFV observables is actually much longer, including for example decays of the $Z$~\cite{Illana:2000ic} and Higgs~\cite{Pilaftsis:1992st,Arganda:2004bz,Arganda:2014dta} bosons, although their current sensitivities for non-unitarity effects are lower\footnote{Nevertheless, a future Tera-Z factory would obtain competitive bounds from LFV $Z$ decays~\cite{Abada:2014cca,DeRomeri:2016gum}.}.

\begin{table}[t!]
\centering
    \begin{tabular}{|c|c|}
         \hline &\\[-2ex]
        Observable & Experimental bound \\[0.5ex]
        \hline\hline
        &\\ [-2ex]
        $\mu\rightarrow e\gamma$
        &$4.2\cdot 10^{-13}$~\cite{MEG:2016leq}\\[1.2ex]
        $\tau\rightarrow e\gamma$
        &$3.3\cdot 10^{-8}$~\cite{BaBar:2009hkt}\\[1.2ex]
        $\tau\rightarrow \mu\gamma$
        &$4.2\cdot 10^{-8}$~\cite{Belle:2021ysv}\\[1.2ex]
        \hline\hline 
        &\\ [-2ex]
        $\mu\rightarrow eee$
        &$1.0\cdot 10^{-12}$~\cite{SINDRUM:1987nra}\\[1.2ex]
        $\tau\rightarrow eee$
        &$2.7\cdot 10^{-8}$~\cite{Hayasaka:2010np}\\[1.2ex]
        $\tau\rightarrow \mu\mu\mu$
        &$2.1\cdot 10^{-8}$~\cite{Hayasaka:2010np}\\[1.2ex]
        \hline\hline
        &\\ [-2ex]
        $\mu\rightarrow e$ (Ti)
        &$4.3\cdot 10^{-12}$~\cite{SINDRUMII:1993gxf}\\[1.2ex]
        $\mu\rightarrow e$ (Au)
        &$7.0\cdot 10^{-13}$~\cite{SINDRUMII:2006dvw}\\[1.2ex]
        \hline
    \end{tabular}
    \caption{Summary of present 90\%CL upper limits for the branching ratios of the most important cLFV observables constraining off-diagonal elements of the 
    matrix $\eta$.
    }
    \label{table:LFV}
\end{table}

Radiative $\ell_\alpha\to\ell_\beta\gamma$ decays are very well-studied and have so far dominated the constraints set by cLFV processes. Their complete rates induced by heavy neutrinos were first reported in Refs.~\cite{Petcov:1976ff,Bilenky:1977du,Cheng:1977vn,Marciano:1977wx,Lee:1977qz,Lee:1977tib}.
In the limit of heavy neutrinos, {\it i.e.}~heavier than $M_W$, and when their potential mass differences can be neglected\footnote{In principle, non-degenerated heavy neutrinos could lead to numerical cancellations in some of the cLFV processes~\cite{Forero:2011pc}. Nevertheless, we consider such a situation very unlikely, as we explain in App.~\ref{App:Cancelations}, and therefore do not take into account in our global analysis.}, these rates can be approximated by
\begin{equation}\label{eq:LFVrad}
{\rm BR}(\ell_\alpha\to \ell_\beta\gamma)\simeq \frac{3\alpha}{2\pi}\,\big|\eta_{\alpha\beta}\big|^2\,,
\end{equation}
and therefore they are directly related to the off-diagonal entries of the non-unitarity matrix $\eta$. 
Notice moreover that this relation is independent of the heavy neutrino mass, as long as it is heavy enough, implying that the radiative decays impose mass-independent bounds on $\eta_{\alpha\beta}$.

On the other hand, the relation between $\eta$ and other cLFV transitions is in general more involved. 
The reason is that, contrary to the radiative decays, other cLFV observables also get contributions from neutral currents from the $Z$ and $H$ boson penguins, as well as from box-diagrams, which have a different dependence on heavy neutrino mixings at large masses~\cite{Alonso:2012ji,Arganda:2016zvc,Herrero:2018luu,Marcano:2019rmk,Calderon:2022alb}, and thus a different $\eta$-dependence. 

As an example to illustrate this behavior, let us consider a minimal scenario with two almost degenerated neutrinos forming a single pseudo-Dirac pair of mass $M$ (see section~\ref{Sec:Results2N}).
Then, the rate for $\mu-e$ conversion in nuclei~\cite{Alonso:2012ji} can be approximated by
\begin{align}\label{eq:LFVmueConv}
{\rm CR}(\mu\to e)\simeq&
\frac{\alpha^5 m_\mu^5G_F^2\,F_p^2}{10368\pi^4 \sw^2\Gamma_{\text{capt}}}\dfrac{Z_{\text{eff}}^4}{Z}\,\big|\eta_{e\mu}\big|^2\, 
\Big|\left(A+Z\right)F_u+\left(2A-Z\right)F_d  \Big|^2\,,
\end{align}
with
\begin{align}
    F_u&=-27-148\sw^2-(27-64\sw^2)\log \frac{M^2}{M_W^2}-(18-48\sw^2)\frac{M^2}{M_W^2}\Tr{\eta},\\
    F_d&=-27+\phantom{1} 74\sw^2+(27-32\sw^2)\log \frac{M^2}{M_W^2}+(18-24\sw^2)\frac{M^2}{M_W^2}\Tr{\eta},
\label{obs:loop_functions}
\end{align}
and $A, Z (Z_{\rm eff}), F_p$ and $\Gamma_{\rm capt}$ are properties of the nucleus: mass number, (effective) atomic number, nuclear form factor and capture rate, respectively.
Notice that we chose $\mu-e$ conversion rate as an example, but similar expressions can be derived for other cLFV processes such as $\mu\to eee$ or $\tau\to eee$.
 
Besides the mild logarithmic dependence, there is a new $M$-dependent contribution, not present in the current with photons and thus in $\ell_\alpha\to\ell_\beta\gamma$.
This new term is suppressed by an additional power of the small $\eta$-matrix, encoded as $\Tr\eta$ in this simplified scenario, but at the same time it is enhanced by the new mass scale, so it can still be important for sufficiently heavy neutrinos. This also happens for other cLFV processes such as the three-body leptonic decays, with additional $M$-dependent terms involving also the diagonal entries of $\eta$.
Consequently, processes such $\mu\to eee$ and $\mu-e$ conversion in nuclei do not impose mass-independent bounds on $\abs{\eta_{e\mu}}$ in the $M\gg M_W$ limit.
Moreover, the experimental limits for these processes cannot be straightforwardly translated to bounds for $\abs{\eta_{e\mu}}$, given the additional dependence on $\Tr{\eta}$. 
Thus, in all generality, the $\abs{\eta_{e\mu}}$ bounds depend not only on the mass but also on the diagonal entries of $\eta$.

\begin{figure}[t!]
\begin{center}
\includegraphics[width=0.95\textwidth]{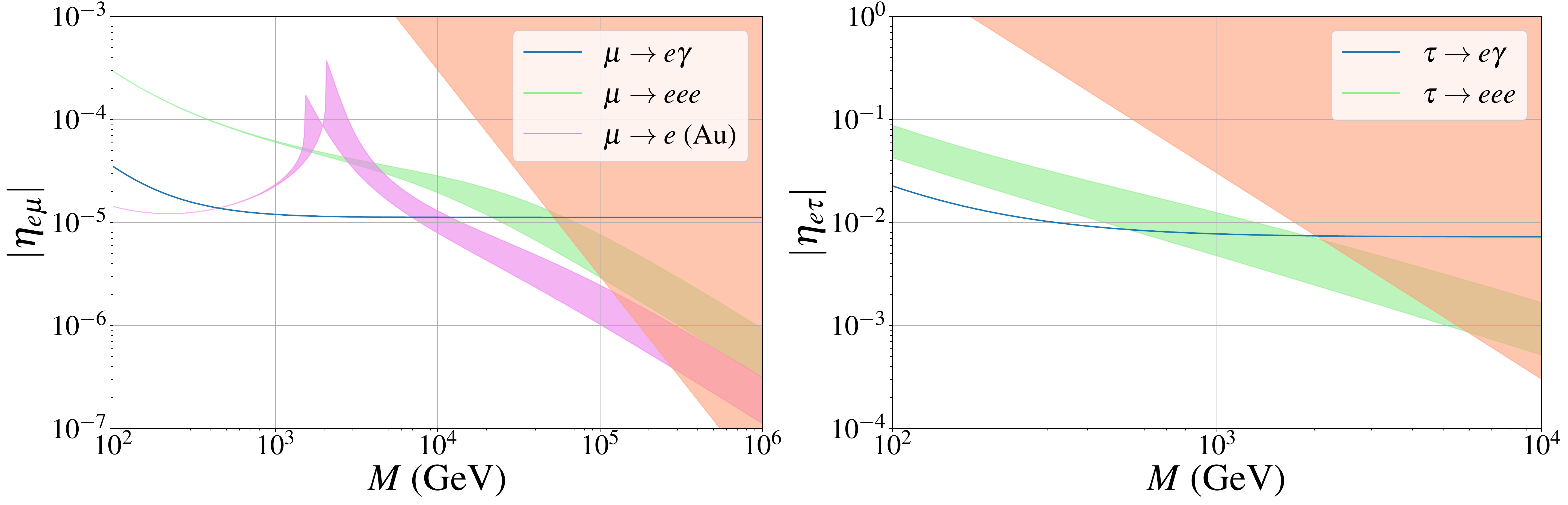}
\caption{
Upper bounds on $\abs{\eta_{e\mu}}$ ($\abs{\eta_{e\tau}}$) imposed by current limits of $\mu\to e\gamma$ ($\tau\to e\gamma$), $\mu\to eee$ ($\tau\to eee$) and $\mu-e$ conversion in gold (see Table~\ref{table:LFV}).
We considered the minimal scenario described in section~\ref{Sec:Results2N} with a normal ordering for light neutrino masses. 
The color bands reflect how the upper bound on $\abs{\eta_{\alpha\beta}}$ is modified as we vary the other free parameters. 
The upper right area in red corresponds to Yukawa couplings larger than 1, outside the perturbativity region.}
\label{fig:cLFVbounds} 
\end{center}
\end{figure}

In order to understand better and to quantify these mass and  $\eta$ dependencies, we show in Fig.~\ref{fig:cLFVbounds} the upper bounds that can be set on $\abs{\eta_{e\mu}}$ and $\abs{\eta_{e\tau}}$ from each of the cLFV rates individually.
This figure corresponds to the most minimal scenario with only a degenerate pair of heavy neutrinos, which will be introduced and explored in detail in section~\ref{Sec:Results2N}, but it is sufficient to exemplify and discuss the overall behavior.
Besides varying the heavy neutrino mass, we also survey 
the values of the additional free parameters of the model (see Section~\ref{Sec:Results2N}) in order to also scan $\Tr{\eta}$ from its minimum\footnote{Notice that in general $\Tr\eta$ cannot be zero for $\eta_{\alpha\beta}>0$ due to the Schwarz inequality in Eq.~\eqref{eq:schwarz}. For example, in the minimal model of section~\ref{Sec:Results2N}, we have $\Tr\eta \geq 2\abs{\eta_{e\mu}}$.} to its maximum value allowed in this scenario. 
This defines a color band for each observable, showing how much the bound on $\eta_{\alpha\beta}$ changes as we vary $\Tr\eta$. 
Since the radiative decays do not depend on $\Tr\eta$, see Eq.~\eqref{eq:LFVrad}, the band is only a line and it can be seen that they quickly saturate to a constant value as soon as the heavy neutrino mass becomes a few times heavier than $M_W$.

On the other hand, the rest of observables display a more complex behaviour. 
Their bounds lie in a band, thicker for heavier masses since the importance of the $\Tr\eta$ terms grows with $M$, and they become more stringent for heavier neutrino masses\footnote{With the exception of a cancellation in the $\mu-e$ conversion rate in gold for masses around 20~TeV. The exact position of this cancellation depends however on the mass and atomic number of the nucleus, so it will be different for each nuclei.}, even overcoming those from the radiative decays.
Nevertheless, whether this crossing between observables happens within the perturbative region of the model and where exactly depend on the values of $\Tr\eta$.
In this particular example, $\mu-e$ conversion dominates over $\mu\to e\gamma$ for masses above 10~TeV approximately, although the exact position varies with $\Tr\eta$.
The situation is similar in the $\tau$-$e$ sector, where $\tau\to eee$ can dominate over $\tau\to e\gamma$ for heavy masses above the TeV range, again depending on the value of $\Tr\eta$.
Notice that in the $\tau$-$e$ sector this crossing happens for lower masses than in the $\mu$-$e$ sector, since in the former we are probing larger values of $\eta$ and, thus, the additional $\eta$-terms are less suppressed. 
Nevertheless, in both sectors these effects become relevant only close to the perturbativity limit, since they require large masses and mixings simultaneously.

For these reasons, which cLFV observable provides the most stringent bound is model-dependent and also changes in different regions of the parameter space. To deal with this fact and to provide the most robust constraints possible, we will ``marginalize over'' the dependencies shown in Fig.~\ref{fig:cLFVbounds} by selecting a bound that would apply to all values of the mass and $\Tr\eta$. In practice we simply consider the constraint stemming from radiative decays as given in Eq.~\eqref{eq:LFVrad}. This choice slightly overestimates the bounds from radiative decays for masses close to $M_W$, where the GIM cancellation starts to be recovered. Nevertheless, in the $\mu$-$e$ sector this effect is compensated by $\mu-e$ conversion in gold, which provides a strikingly similar constraint precisely in that region of the parameter space. While this is not the case in the $\tau$ sector, those cLFV bounds are subdominant compared to those derived from the Schwarz inequality, so that its effect is not relevant in the fit. Consequently, Eq.~\eqref{eq:LFVrad} turns out to be a good approach to a conservative and mass-independent cLFV limit, and will be added to our fit. In any event, it should be noted that in some regions of the parameter space, for the heaviest masses, stronger constraints than our conservative estimate might apply as shown Fig.~\ref{fig:cLFVbounds}.

\section{Global fit bounds for the 2 neutrino case (2N-SS)}
\label{Sec:Results2N}

In order to reproduce the two distinct mass splittings that characterize the neutrino oscillation phenomenon, at least two of the mainly-SM neutrinos need to become massive. Therefore the minimum number of heavy neutrinos needed in order to have a realistic neutrino mass model is two, as we consider in this section. 
We will dub this setup with two heavy neutrinos as the \textbf{2N-SS}, with RH neutrinos $N_R$ and $N'_R$ inducing the observed light neutrino masses and mixings.

Moreover, in order for these RH neutrinos to have sizeable mixing with the active neutrinos while generating radiatively stable and small neutrino masses, a lepton number protected seesaw realization is required with the heavy neutrinos forming a pseudo-Dirac pair~\cite{Kersten:2007vk,Abada:2007ux,Gavela:2009cd,Moffat:2017feq}.
The most general neutrino mass matrix which satisfies these characteristics, in the basis $\left(\nu_L \hspace{0.15cm}N^c_R\hspace{0.15cm}{N'}_R^c \right)^T$, has the following form:
\begin{equation}
	\mathcal{M}_\nu=
	\begin{pmatrix}
		0&Yv/\sqrt{2}&\epsilon Y'v/\sqrt{2}\\
		Y^Tv/\sqrt{2}&\mu'&M\\
		\epsilon{Y'}^Tv/\sqrt{2}&M&\mu
	\end{pmatrix},
\label{2NSS:massmatrix}
\end{equation}
where the $\epsilon, \mu$ and $\mu'$ terms softly break a generalized lepton number symmetry, $L_e=L_\mu=L_\tau=L_N=-L_{N'}=1$. Therefore, it is technically natural to consider $\epsilon,\mu/M,\mu'/M\ll1$. 
Indeed, in the limit when these terms vanish, $\mathcal{M}_\nu$ preserves the symmetry, yielding three massless neutrinos ($m_\nu=0$) and two degenerate heavy neutrinos forming a Dirac fermion. 
However, even in this limit the active-heavy mixings are already non-zero and hence unsuppressed by the smallness of neutrino masses
\begin{equation}\label{eq:mixing2N}
	\Theta_\alpha=\begin{pmatrix}
		0&\dfrac{Y_\alpha v}{\sqrt{2}M}
	\end{pmatrix}
\equiv\begin{pmatrix}
	0&\theta_\alpha
\end{pmatrix},
\end{equation}
leading to potentially sizable non-unitarity effects given by
\begin{equation}
	\eta=\dfrac{1}{2}\begin{pmatrix}
		\abs{\theta_e}^2&\theta_e\theta_\mu^*&\theta_e\theta_\tau^*\\
		\theta_e^*\theta_\mu&\abs{\theta_\mu}^2&\theta_\mu \theta_\tau^*\\
		\theta_e^*\theta_\tau&\theta_\mu^*\theta_\tau&\abs{\theta_\tau}^2
	\end{pmatrix}.
\label{2NSS:eta}
\end{equation}

The minimality of this model has two important implications for our analysis. 
On the one hand, this $\eta$-matrix saturates the Schwarz inequality ({\it i.e.}~$\abs{\eta_{\alpha\beta}}=\sqrt{\eta_{\alpha\alpha}\eta_{\beta\beta}}$), meaning that the cLFV bounds will constrain not only the off-diagonal entries of $\eta$, but also the diagonal ones. 
Therefore the cLFV observables need to be added into the global fit together with the LFC ones in Table~\ref{table:observables}.
On the other hand, the flavor structure of the mixing is subject to important constraints from the requirement of explaining the correct light neutrino mass matrix. In fact, this flavor structure can be reconstructed from the light neutrino masses and mixings \cite{Gavela:2009cd}, except for an overall scale $\theta$:
\begin{align}
	\theta_\alpha&=\dfrac{\theta}{\sqrt{2}}\left(\sqrt{1+\rho}\hspace{0.1cm} U^*_{\alpha 3}+ \sqrt{1-\rho}\hspace{0.1cm}U^*_{\alpha2}\right)&&\text{for Normal Ordering (NO),} \label{eq:2Nparam_NO}\\
	\theta_\alpha&=\dfrac{\theta}{\sqrt{2}}\left(\sqrt{1+\rho}\hspace{0.1cm} U^*_{\alpha 2}+ \sqrt{1-\rho}\hspace{0.1cm}U^*_{\alpha1}\right)&&\text{for Inverted Ordering (IO),}\label{eq:2Nparam_IO}
\end{align}
where
\begin{align}
    \rho &= \dfrac{\sqrt{\Delta m^2_{31}}-\sqrt{\Delta m^2_{21}}}{\sqrt{\Delta m^2_{31}}+\sqrt{\Delta m^2_{21}}}&&\text{for NO,}
    \\
    \rho &= \dfrac{\sqrt{\Delta m^2_{23}}-\sqrt{\Delta m^2_{23}-\Delta m^2_{21}}}{\sqrt{\Delta m^2_{23}}+\sqrt{\Delta m^2_{23}-\Delta m^2_{21}}}&&\text{for IO,}
\end{align}
with $\Delta m^2_{ij}=m^2_i-m^2_j$ and where we use for $U$ the standard PDG parametrization with a single Majorana phase, as one neutrino remains massless. 

From oscillation experiments the values of the two light neutrino mass splittings and the three mixing angles have been determined with good accuracy. However, the Dirac CP phase and Majorana phases are still mostly unconstrained. As such, in our analysis we will fix the mass splittings and the mixing angles to their best-fit values as reported by NuFIT~\cite{Esteban:2020cvm}, while treating the Dirac phase $\delta$, the Majorana phase $\phi$ and the overall magnitude\footnote{Actually $\theta$ is a complex parameter. However, only its modulus is relevant for our analysis, as its phase only appears in lepton number violating processes, which are not studied in this work.} of the mixing $\theta$ as free parameters for our fit. All in all, this scenario is fully described by 3 free parameters.

\begin{figure}[t!]
    \centering
    \includegraphics[width=0.95\textwidth]{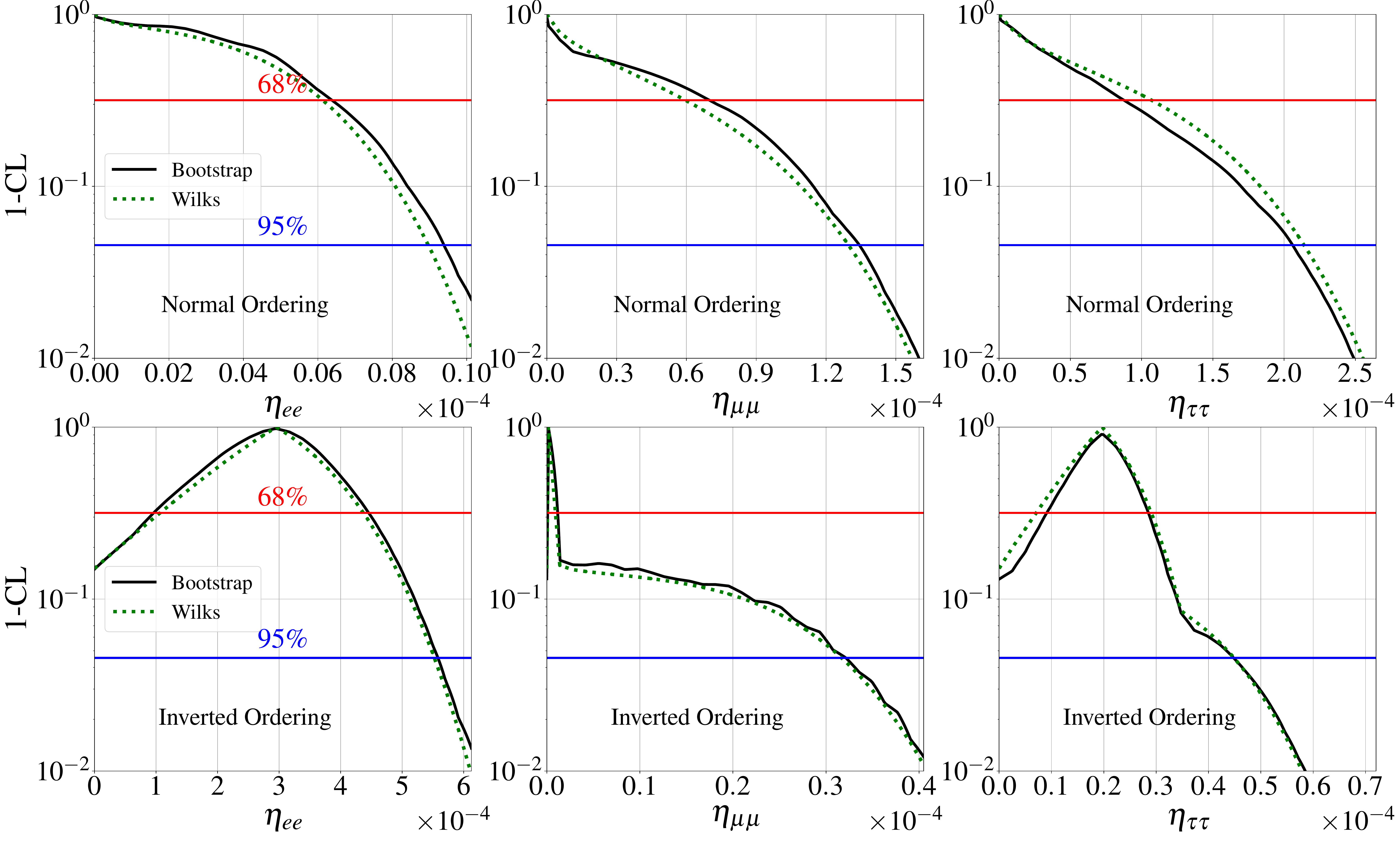}
    \caption{  
    Profiles from our global fit analysis of the minimal model with 2 RH neutrinos (2N-SS), for NO (upper panels) and IO (lower panels). 
    In each panel, we minimized over all the parameters of the fit but the corresponding $\eta_{\alpha\alpha}$.
    The dashed green lines are obtained following Wilks' theorem, while the solid black lines are the result of calibrating the test statistic by bootstrapping it (see Appendix~\ref{App:Procedure}).}
    \label{fig:2NSSprofiles}
\end{figure}

The results of our statistical analysis for this 2N-SS setup are shown in Fig.~\ref{fig:2NSSprofiles}, where we plot the profiles for the diagonal entries of the $\eta$-matrix for the two neutrino mass orderings.
In each panel, the dashed green line is obtained following Wilks' theorem, while the solid black line is the result of calibrating our test statistic through the  bootstrapping procedure (see App.~\ref{App:Procedure}), since, given the strong correlations among the observables implied by Eqs.~\eqref{eq:2Nparam_NO} and~\eqref{eq:2Nparam_IO}, the requirements for Wilks' theorem to apply are not met. Nevertheless, we find that in this case Wilks' theorem still provides a good approximation. Indeed, given the little freedom available to this rather constrained scenario, the bound on the non-unitarity parameter $\theta$ mainly stems from the best constrained observable in the fit, namely $\mu \to e \gamma$, and, neglecting the contributions of the other observables, the requirements for Wilks' theorem to apply are approximately met. However, we will see that this is not the case for other, less constrained, scenarios. 

\begin{table}[t!]
    \centering
    \begin{tabular}{|c|c|c||c|c|}
    \hline 
    &\multicolumn{2}{c||}{}&\multicolumn{2}{c|}{}\\[-2.5ex]
    \multirow{2}{*}{\bf 2N-SS}&\multicolumn{2}{c||}{Normal Ordering} 
    &\multicolumn{2}{c|}{Inverted Ordering}\\[0ex]  
    \cline{2-5} &&&&\\[-2.5ex]
    & $68\%$CL& $95\%$CL& $68\%$CL& $95\%$CL\\
    \hline
    &&&&\\[-2ex]
    $\etaee=\dfrac{\thetae^2}{2}$ & $6.4\cdot10^{-6}$ & $9.4\cdot10^{-6}$ & $[0.98,4.4]\cdot10^{-4}$ & $5.5\cdot10^{-4}$ \\[1.2ex]
    $\etamumu=\dfrac{\thetamu^2}{2}$ & $6.9\cdot10^{-5}$ & $1.3\cdot10^{-4}$  &$[0.20,1.0]\cdot10^{-6}$ & $3.2\cdot10^{-5}$\\[1.2ex] 
    $\etatautau=\dfrac{\thetatau^2}{2}$ &$8.6\cdot10^{-5}$&$2.1\cdot10^{-4}$&$[0.94,2.8]\cdot10^{-5}$&$4.5\cdot10^{-5}$ \\[1.2ex]
    $\Tr{\eta}=\dfrac{\abs{\theta}^2}{2}$ & $1.6\cdot10^{-4}$ & $2.9\cdot10^{-4}$ & $[1.1,4.8]\cdot10^{-4}$ & $6.0\cdot10^{-4}$ \\[1.2ex]
    $\abs{\etaemu}=\dfrac{\abs{\theta_e\theta_\mu^*}}{2}$ &$8.3\cdot10^{-6}$ &$1.2\cdot10^{-5}$ & $[0.37,1.0]\cdot10^{-5}$ & $1.3\cdot10^{-5}$\\[1.2ex]
    $\abs{\etaetau}=\dfrac{\abs{\theta_e\theta_\tau^*}}{2}$ & $1.5\cdot10^{-5}$ &$2.2\cdot10^{-5}$ & $[0.25,1.2]\cdot10^{-4}$ & $1.4\cdot10^{-4}$\\[1.2ex]
    $\abs{\etamutau}=\dfrac{\abs{\theta_\mu\theta_\tau^*}}{2}$ &$7.2\cdot10^{-5}$ &$1.3\cdot10^{-4}$ & $[0.38,3.0]\cdot10^{-6}$ & $3.5\cdot10^{-5}$\\[1.2ex]
    
    \hline    
    \end{tabular}
    \caption{Upper bounds (or preferred intervals) for the most minimal set-up with two heavy neutrinos forming a single pseudo-Dirac pair (2N-SS), which are obtained from the boostrapped profiles in Fig.~\ref{fig:2NSSprofiles} and the equivalent ones (not-shown) for $\Tr\eta$ and the off-diagonal elements.
    Note that these results directly apply to $\eta$ and to (half of) the squared active-sterile mixings $|\theta_\alpha|^2$. They can also be easily translated to the $\alpha$-parametrization, as detailed in the text.
    }
    \label{table:2NSSbounds}
\end{table}

The resulting 68\% and 95\%CL upper limits (or preferred intervals in some cases) are summarised in Table~\ref{table:2NSSbounds}. Here and in all the RH neutrino scenarios under study, we also provide the constraints derived for $\Tr\eta$, which in this simplest case amounts to $\theta^2/2$, as a measurement of how large the \emph{total} deviation from unitarity of the whole matrix is allowed to be regardless of its particular flavour structure. Indeed, $\Tr\eta$ is an invariant under changes of basis and, given that $\eta$ is positive definite by construction, its trace corresponds to the sum of its three eigenvalues. 

It is interesting to note that in this scenario the results are rather different between the profiles for normal and inverted orderings.
Besides the different ranges for each $\eta_{\alpha\alpha}$, as it can be seen in Fig.~\ref{fig:2NSSprofiles}, we find a non-unitary best-fit point for IO, while this does not happen for NO.
This qualitative difference, which also conditions the different values obtained for the allowed ranges, can be understood as an interplay between the preference of the data and the constrained flavor structure of $\eta$ in this model. 

\begin{figure}[t!]
\centering
\includegraphics[width=.9\textwidth]{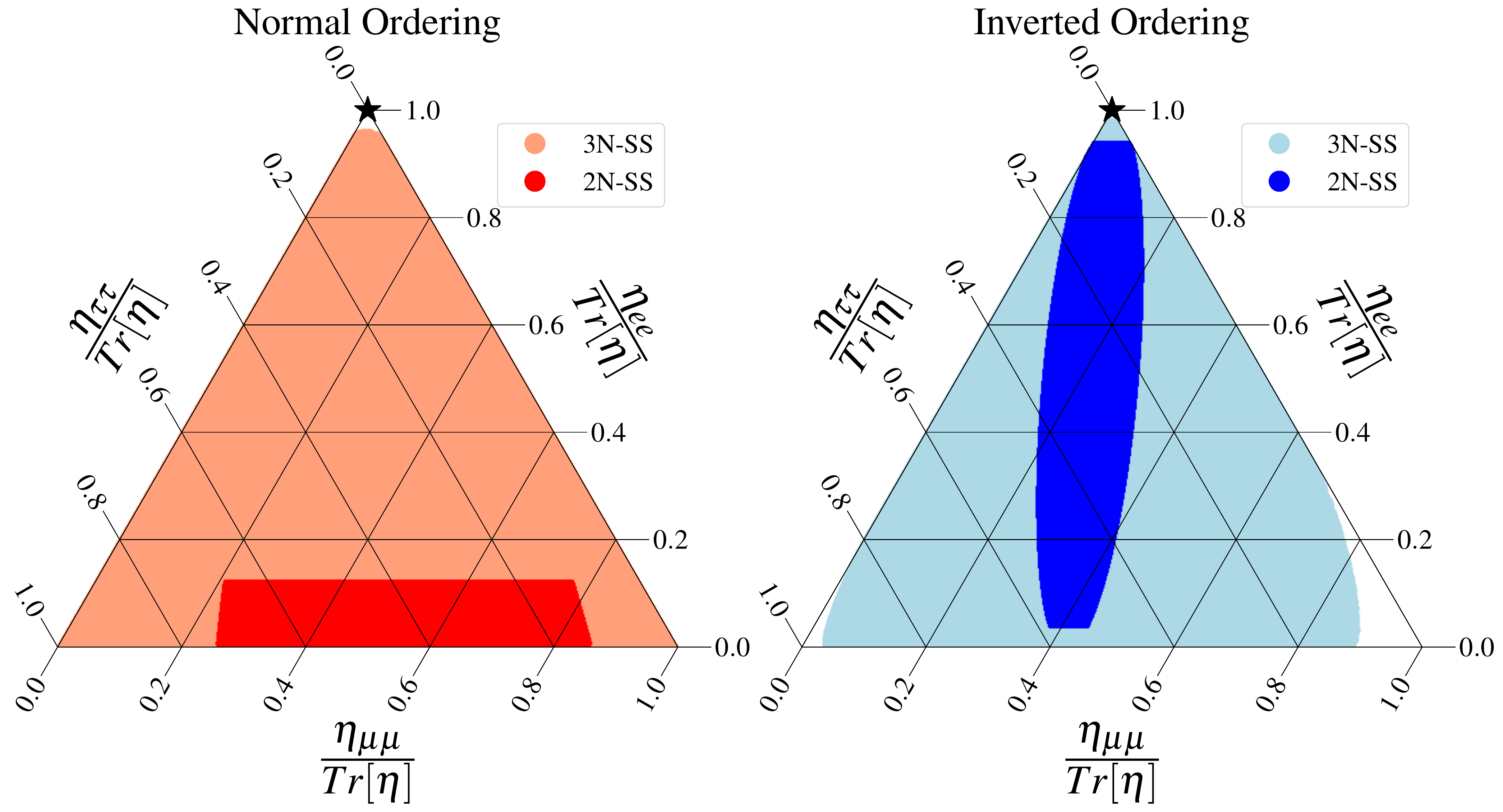}
\caption{Regions in the RH neutrino mixing flavor space  consistent with current neutrino oscillation data, fixing the mixing angles to their best fit values, and varying the phases and absolute neutrino mass. The darker regions correspond to the minimal scenario with 2 RH neutrinos (2N-SS)~\cite{Caputo:2017pit} described in section~\ref{Sec:Results2N}, while the lighter regions are for the next-to-minimal scenario with 3 RH neutrinos (3N-SS)~\cite{Chrzaszcz:2019inj} studied in section~\ref{Sec:Results3N}, where we include the cosmological upper bound on neutrino masses.
The black star corresponds to the best-fit point for the general neutrino case (G-SS, see section~\ref{Sec:ResultsGSS}), where the hole triangle is allowed, indicating the actual preference of the data set given in Table~\ref{table:observables}. Note that in the 2N-SS and 3N-SS, the axes correspond to the normalized squared mixings since $\eta_{\alpha\alpha}/\Tr\eta=|\theta_\alpha|^2/|\theta|^2$.
}\label{fig:triangles}
\end{figure}

On the one hand, in Table~\ref{table:observables} a mild ($\sim1\sigma$) preference for a non-zero $\etaee+\etamumu$ from $M_W$ and $\sw^2$ can be seen, as well as an also mild ($\sim1\sigma$) preference for $\etaee>\etamumu$ from the LFU ratios. 
Furthermore, the CKM data strongly disfavors non-zero values of $\etamumu$, as they can only worsen the Cabibbo anomaly, and there is also no preference for a non-zero $\etatautau$.
On top of that, the strong cLFV bounds\footnote{
Notice that in the 2N-SS we expect mixings to all three flavors, since all of them are proportional to the overall scale $\theta$.
This implies that the cLFV rates cannot be avoided by turning just a single $\theta_\alpha$ off, and thus they are specially constraining for this 2N-SS scenario.}
on the $\mu$-$e$ sector require either the electron or the muon mixing to be very small.
Combining all these aspects, we obtain that the data prefers a non-zero $\etaee$ with suppressed $\etamumu$ and $\etatautau$.
This is represented with a black star in Fig.~\ref{fig:triangles}, where we display the possible flavor patterns for both normal and inverted orderings and for the different scenarios under consideration.

On the other hand, the flavor structure of the 2N-SS is very restricted.
It is determined up to the mass ordering and the unknown phases $\delta$ and $\phi$, and the resulting regions in flavor space are shown as darker areas in Fig.~\ref{fig:triangles}.
In particular, the NO case is characterised by having a suppressed $\thetae$ with respect to $\thetamu$ and $\thetatau$ (for all values of $\delta$ and $\phi$), which is precisely the opposite of what the data prefers.
This is manifest in the figure, where the dark red region is far away from the data-preferred black star.
Therefore, the NO has the best fit-point at $\eta=0$, with stronger bounds for $\etaee$ than for the other flavors. 
On the contrary, the flavor structure of the IO case is such that, for certain values of the phases, the mixing to the electron overcomes that of the muon and tau, and thus it can accommodate better the preference of the data.
We see it again from Fig.~\ref{fig:triangles}, as the dark blue region approaches more to the black star.
Consequently, the IO has a non-trivial best-fit point, with a mild ($\sim1\sigma$) preference for non-zero $\eta$, where the best-fit for $\etamumu$ is suppressed with respect to $\etaee$ and $\etatautau$.
Consistently, we find a $\Delta \chi^2 = 2.07$ in favour of the IO minimum with respect to the NO one.

Regarding the off-diagonal elements, the bound on $\abs{\etaemu}$ is completely dominated by the corresponding cLFV bound. Conversely, the bounds on $\abs{\etaetau}$ and $\abs{\etamutau}$ are much stronger than the bounds derived from their corresponding cLFV processes. This effect is related to the saturation of the Schwarz inequality in Eq.~\eqref{eq:schwarz} and the strong correlations present in the flavor structure of the 2N-SS: since all three mixings are proportional to a common scale, this also implies $\abs{\etaemu}\propto\abs{\etaetau}\propto\abs{\etamutau}$ where the proportionality depends on the phases $\delta$ and $\phi$. Thus, the very strong bound on $\abs{\etaemu}$ also induces quite stringent bounds on $\abs{\etaetau}$ and $\abs{\etamutau}$.

It should be noted that, even though the Schwarz inequality is saturated and one can reconstruct the off-diagonal elements from the diagonal ones, the bounds for the former cannot be inferred from the latter. This is a consequence of the strong correlations imposed by the cLFV constraints and of the fact that, in order to obtain each bound, the rest of the parameters are profiled over.
Therefore, it is not possible to saturate simultaneously all bounds and it should be checked that all constraints are satisfied for a given mixing pattern.

As an alternative to $\eta$, a lower triangular parametrization~\cite{Xing:2007zj,Escrihuela:2015wra} has been shown to be more convenient for the study of non-unitarity in the neutrino oscillation phenomenon~\cite{Blennow:2016jkn}:
\begin{equation}
 \alpha=\begin{pmatrix}
  \alpha_{11}&0&0\\
 \alpha_{21 }&\alpha_{22}&0\\
\alpha_{31}&\alpha_{32}&\alpha_{33}
    \end{pmatrix}.
\end{equation}
These $\alpha$ parameters can be straightforwardly mapped to the $\eta$-matrix~\cite{Blennow:2016jkn}, thus our bounds can be easily translated to this parametrization.
In particular, at $95\%$CL we obtain:
\begin{align}
\text{NO:}&\hspace{0.5cm}    \abs{\mathbb{I}-\alpha}<\begin{pmatrix}
        9.4\cdot10^{-6}&0&0\\
        2.4\cdot10^{-5}&1.3\cdot10^{-4}&0\\
        4.4\cdot10^{-5}&2.6\cdot10^{-4}&2.1\cdot10^{-4}
    \end{pmatrix}
    \,,\\
\text{IO:}&\hspace{0.5cm}    \abs{\mathbb{I}-\alpha}<\begin{pmatrix}
        5.5\cdot10^{-4}&0&0\\
        2.6\cdot10^{-5}&3.2\cdot10^{-5}&0\\
        2.8\cdot10^{-4}&7.0\cdot10^{-5}&4.5\cdot10^{-5}
    \end{pmatrix}
    .
\end{align}

\section{Global fit bounds for the 3 neutrino case (3N-SS)}
\label{Sec:Results3N}

The scenario in the previous section with two heavy neutrinos is the most minimal set-up to accommodate oscillation data, however it predicts the lightest neutrino to be massless. 
While this is in agreement with current observations, the overall scale of neutrino masses still remains unknown and, thus, it is perfectly possible that all light neutrinos are massive. 
In that case, the minimum number of extra neutrinos needed to accommodate light neutrino masses is three, in line with the three SM generations for all other fermions. In this section we will investigate the bounds that may be derived in this scenario that we dub \textbf{3N-SS}. 

Similarly to the previous section, the only way in which sizeable mixing can be obtained while keeping neutrino masses small and stable under radiative corrections is via a lepton number protected Lagrangian. More precisely, in the basis $\left(\nu_L \hspace{0.15cm}N^c_R\hspace{0.15cm}{N'}_R^c \hspace{0.15cm}{N''}_R^c\right)^T$, the mass matrix reads:
\begin{equation}
	\mathcal{M}_\nu=\begin{pmatrix}
		0&Yv/\sqrt{2}&\epsilon_1 Y'v/\sqrt{2}&\epsilon_2 Y'' v/\sqrt{2}\\
		Y^Tv/\sqrt{2}&\mu_1&M&\mu_3\\
		\epsilon_1{Y'}^Tv/\sqrt{2}&M&\mu_2&\mu_4\\
		\epsilon_2{Y''}^T v/\sqrt{2}&\mu_3&\mu_4&M'
	\end{pmatrix},
	\label{3NSS:massmatrix}
\end{equation}
where $\epsilon_1,\epsilon_2\ll1$ and the $\mu_i$ parameters are small compared with $M,M'$. 
Assigning $L_e=L_\mu=L_\tau=L_N=-L_{N'}=1$ and $L_{N''}=0$, in the lepton number conserving limit two heavy neutrinos arrange into a Dirac fermion with non-vanishing mixing with the active neutrinos, while the third heavy neutrino behaves as a decoupled Majorana state. Therefore, the structure of the $\eta$ matrix is exactly the same as in  Eq.~\eqref{2NSS:eta} and thus the Schwarz inequality is again saturated, which requires the inclusion of cLFV bounds in combination with the LFC observables in our global fit.

However, the relations between the mixings to the active flavors are different from the ones present in the 2N-SS, as the increased freedom in the parameter space implies different (looser) correlations imposed by correctly reproducing the observed neutrino mass matrix. 
In particular, it has been shown in~\cite{Fernandez-Martinez:2015hxa} that the mixing to one flavor is determined by the other two and the entries of the light neutrino mass matrix. For instance, if one chooses to reconstruct the $\tau$ mixing, then:
\begin{align}
		\theta_\tau\simeq \frac{1}{m_{e\mu}^2-m_{ee}m_{\mu\mu}}&\Biggl(\theta_e\left(m_{e\mu}m_{\mu\tau}-m_{e\tau}m_{\mu\mu}\right)+\theta_\mu (m_{e\mu}m_{e\tau}-m_{\mu\tau}m_{ee}) \nonumber\\
		+& \sqrt{\theta_e^2 m_{\mu\mu}-2\theta_e\theta_\mu m_{e\mu}+\theta_{\mu}^2 m_{ee}}\,\times\nonumber\\
		\times&\sqrt{m_{e\tau}^2m_{\mu\mu}-2m_{e\mu}m_{e\tau}m_{\mu\tau}+m_{ee}m_{\mu\tau}^2+m_{e\mu}^2m_{\tau\tau}-m_{ee}m_{\mu\mu}m_{\tau\tau}} \Biggr),
 \label{3NSS:mixing_correlation}
\end{align}
where $m_{\alpha\beta}\equiv\left(m_\nu\right)_{\alpha\beta}=\left(Um_{\rm diag}\,U^T\right)_{\alpha\beta}$ are the entries of the light neutrino mass matrix.

In our analysis, we fix the mass splittings and the mixing angles to their best-fit values, leaving the Dirac phase $\delta$, the two Majorana phases $\phi_1$ and $\phi_2$, and the lightest neutrino mass $\mlightest$ as the only free parameters in $m_\nu$. Additionally, we have also as free parameters the mixings $\theta_e$ and $\theta_\mu$, both in modulus and phase. In total, this setup is characterized by 8 free parameters. Among these parameters, the only one that is constrained by experimental data is $\mlightest$, which is currently bounded from kinematical searches at KATRIN~\cite{KATRIN:2021uub} and, more strongly, from cosmology~\cite{Planck:2018vyg}.
We provide our results considering the more stringent bound from Planck of $\sum m_\nu<0.12$~eV (95$\%$CL). Nevertheless, we have also performed the analysis using the looser constraint from KATRIN and obtained very similar results. 

\begin{figure}[t!]
    \centering
    \includegraphics[width=0.95\textwidth]{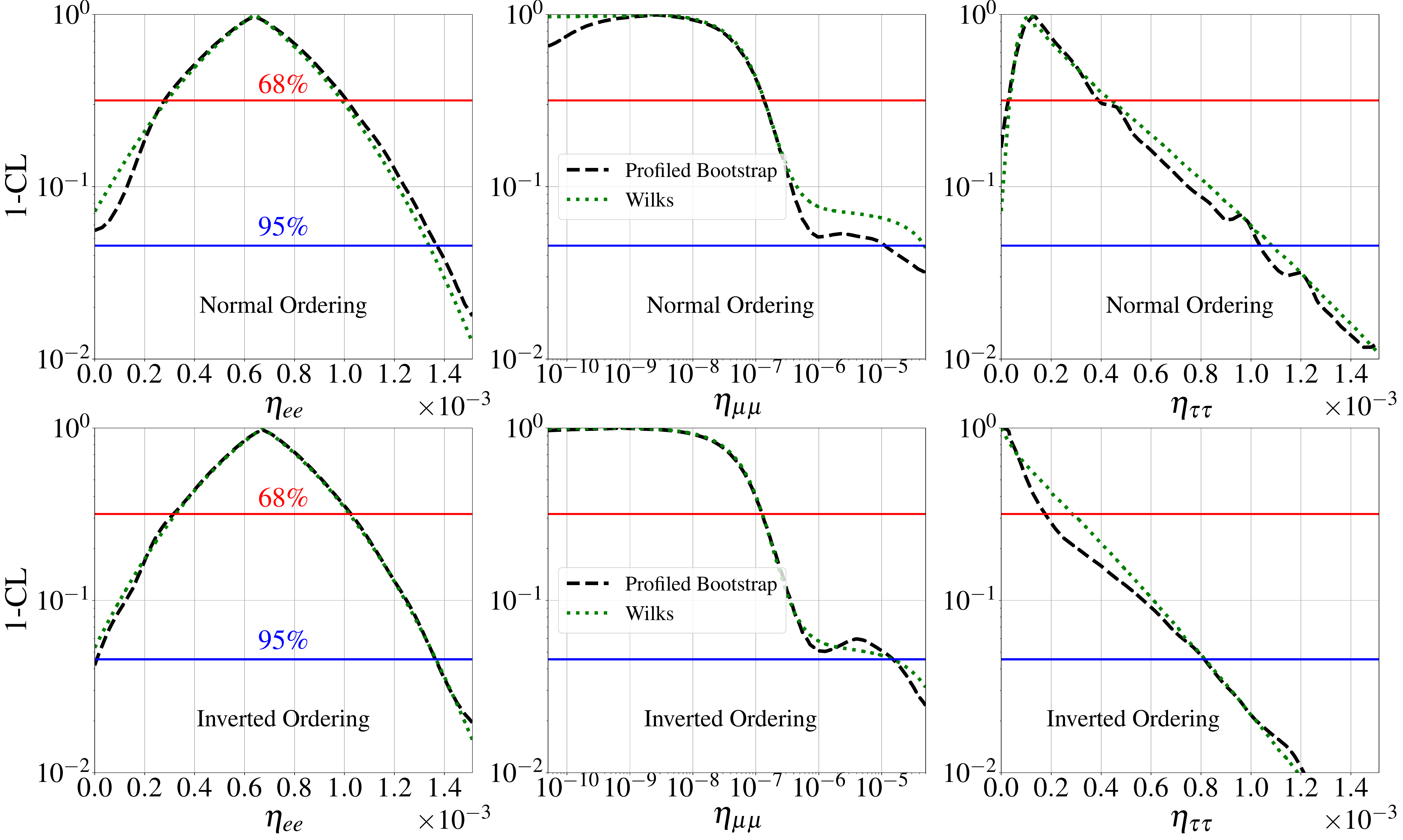}
    \caption{ 
    Same as Fig.~\ref{fig:2NSSprofiles} but for the scenario with 3 RH neutrinos (3N-SS). In this case, however, the calibration of the test statistic is done following the profiled bootstrapping procedure described in Appendix~\ref{App:Procedure}.
    We have used the Planck upper bound~\cite{Planck:2018vyg} on neutrino masses, but similar results are obtained using the KATRIN bound~\cite{KATRIN:2021uub}.
    Notice the different (log)scale for $\eta_{\mu\mu}$.}
    \label{fig:3NSSprofile}
\end{figure}

The results of the analysis are shown in Fig.~\ref{fig:3NSSprofile} and Table~\ref{table:3NSSbounds}.
Given the large dimensionality of the parameter space of this setup, the bootstrapping procedure we followed in the 2N-SS is no longer tractable, as the number of points one has to calibrate grows exponentially with the number of free parameters. As a consequence, we are forced to perform an approximated bootstrap procedure, which we will dub {\it Profiled Bootstrap}, and whose details can be found in Appendix~\ref{App:Procedure}.

\begin{table}[t!]
    \centering
    \begin{tabular}{|c|c|c||c|c|}
    \hline 
    &\multicolumn{2}{c||}{}&\multicolumn{2}{c|}{}\\[-2.5ex]
    \multirow{2}{*}{\bf 3N-SS}&\multicolumn{2}{c||}{Normal Ordering} 
    &\multicolumn{2}{c|}{Inverted Ordering}\\[0ex]  
    \cline{2-5} &&&&\\[-2.5ex]
    & $68\%$CL& $95\%$CL& $68\%$CL& $95\%$CL\\
    \hline
    &&&&\\[-2ex]
    $\etaee=\dfrac{\thetae^2}{2}$ & $[0.28,0.99]\cdot10^{-3}$ & $1.3\cdot10^{-3}$ &$[0.31,1.0]\cdot10^{-3}$ & $1.4\cdot10^{-3}$\\[1.2ex] 
    $\etamumu=\dfrac{\thetamu^2}{2}$ &$1.3\cdot10^{-7}$&$1.1\cdot10^{-5}$&$1.2\cdot10^{-7}$&$1.0\cdot10^{-5}$ \\[1.2ex]
    $\etatautau=\dfrac{\thetatau^2}{2}$ & $[0.3,3.9]\cdot10^{-4}$ & $1.0\cdot10^{-3}$ & $1.7\cdot10^{-4}$ & $8.1\cdot 10^{-4}$ \\[1.2ex]
    $\Tr{\eta}=\dfrac{\abs{\theta}^2}{2}$ & $[0.35,1.3]\cdot10^{-3}$ & $1.9\cdot10^{-3}$ & $[0.33,1.0]\cdot10^{-3}$ & $1.5\cdot10^{-3}$ \\[1.2ex]
    $\abs{\etaemu}=\dfrac{\abs{\theta_e\theta_\mu^*}}{2}$ & $8.5\cdot10^{-6}$ & $1.2\cdot10^{-5}$ & $8.5\cdot10^{-6}$ & $1.2\cdot 10^{-5}$ \\[1.2ex]
    $\abs{\etaetau}=\dfrac{\abs{\theta_e\theta_\tau^*}}{2}$ & $[1.3,5.1]\cdot10^{-4}$ & $9.0\cdot10^{-4}$ & $3.3\cdot10^{-4}$ & $8.0\cdot 10^{-4}$ \\[1.2ex]
    $\abs{\etamutau}=\dfrac{\abs{\theta_\mu\theta_\tau^*}}{2}$ & $5.0\cdot10^{-6}$ & $5.7\cdot10^{-5}$ & $3.8\cdot10^{-6}$ & $1.8\cdot 10^{-5}$ \\[1.2ex]
    
    \hline    
    \end{tabular}
    \caption{Upper bounds (or preferred intervals) for the 3N-SS scenario, considering the cosmological upper bound on neutrino masses. 
    The results for the diagonal entries of $\eta$ are obtained from the {\it profiled bootstrap} in Fig.~\ref{fig:3NSSprofile}, while the off-diagonal ones follow Wilks' theorem, as explained in App.~\ref{App:Procedure}.
    Similar results are obtained when considering instead the KATRIN upper bound.}
    \label{table:3NSSbounds}
\end{table}

We find that assuming Wilks' theorem does not deviate much from the results of the approximate bootstrapping, with the main differences usually appearing close to the physical border at $\eta_{\alpha \alpha} = 0$.
 For the $\eta_{ee}$ profile, we find a non-zero best-fit value and a slight enhancement of the CL with respect to Wilks' near $\eta_{ee}=0$ due to boundary effects expected~\cite{Feldman:1997qc}. For the $\eta_{\mu\mu}$ profile, we find a more complex behaviour. 
 For lower CL, since the data prefers $\eta_{ee}\neq0$, the $\eta_{\mu\mu}$ profile is dominated by the cLFV bound, which leads to a good agreement with Wilks' theorem and to a very stringent $68\%$ bound.
 However, at higher CL, when $\eta_{ee}$ is allowed to vanish by the rest of the observables, $\mu\rightarrow e\gamma$ ceases to be a relevant bound, and $\eta_{\mu\mu}$ is instead constrained by the LFC observables, which are looser. As a result, the profile develops a plateau-like feature at the same CL at which $\eta_{ee}=0$. Furthermore, we observe an enhancement of the CL in the plateau region, with respect to Wilks' expectation coincident with the same enhancement for $\eta_{ee}=0$ which causes the plateau. 

Contrary to the 2N-SS case, the results do not depend significantly on the neutrino mass ordering. 
In particular, we find a non-zero best-fit now for both mass orderings, since the additional freedom with respect to the 2N-SS setup allows for accommodating oscillation data with larger $\eta_{ee}$ also for NO. 
This can be seen in Fig.~\ref{fig:triangles}, where the lighter regions cover a much bigger area of the triangle for both NO and IO.
The only substantial difference is that, for NO, there is a slight preference ($1-2\sigma$) for a non-zero $\eta_{\tau \tau}$. This is a result of the correlation between the mixings present in the 3N-SS. As the best-fit point stands at $\eta_{ee}\neq0$ and $\eta_{\mu\mu}=0$, given the correlation in Eq.~\eqref{3NSS:mixing_correlation} this also induces a best-fit point for $\eta_{\tau \tau}\neq0$ in the NO case. However, this does not happen for IO, since its flavor structure allows for a suppressed $\eta_{\tau \tau}$ even when $\eta_{ee}\neq0$ (see Fig.~\ref{fig:triangles}). Consequently, since the data prefers a suppressed $\eta_{\tau \tau}$, the IO case yields again a slightly better fit with respect to the NO, although less pronounced compared to the 2N-SS scenario.
More precisely, we find a $\Delta\chi^2=0.5$ in favor of the IO best-fit with respect to the NO one.

Regarding the bounds on the off-diagonal entries, it is interesting to compare the global bounds with those imposed directly by cLFV observables\footnote{These bounds correspond to the {\it LFV bound} column in Table~\ref{table:GSSbounds}.} following Eq.~\eqref{eq:LFVrad}.
We see that the constraints on $\abs{\etaemu}$ are very similar to the ones derived directly from cLFV processes, which reflects that in the $\mu$-$e$ sector the test statistics is dominated by the strong bounds on $\mu\to e$ transitions.
In the $e$-$\tau$ sector, we find a mild ($1-2\sigma$) preference for a non-zero $\abs{\etaetau}$, which is induced from the correlation given in Eq.~\eqref{3NSS:mixing_correlation} that implies a non-zero $\thetatau$ for a non-zero $\thetae$.
This effect is softer for IO, since, as explained previously, its flavor structure allows for a suppressed $\thetatau$ even when $\thetae$ is non-zero. 
Nevertheless, when focusing on the $95\%$ C.L. upper limits, we see that the bounds in both the $e$-$\tau$ and $\mu$-$\tau$ sectors are much stronger than those derived from the corresponding cLFV processes. 
This is especially true for $\abs{\eta_{\mu\tau}}$, which  
can be understood in terms of the $\etamumu$ profiles in Fig.~\ref{fig:3NSSprofile}: as previously stated, $\etamumu$ is constrained to be very small up to relatively high CL due to the preference for a non-zero $\etaee$ and the very strong cLFV bound on $\abs{\eta_{e\mu}}=\sqrt{\etaee\etamumu}$. This tight bound on $\etamumu$ also induces a strong bound on $\abs{\etamutau}=\sqrt{\etamumu\etatautau}$. Note that this interplay is a consequence of the saturation of the Schwarz inequality.

Finally, the bounds shown in Table~\ref{table:3NSSbounds} can be again translated to the lower-triangular $\alpha$ parametrization obtaining, at $95\%$CL:
\begin{align}
\text{NO:}&\hspace{0.5cm}
    \abs{\mathbb{I}-\alpha}<\begin{pmatrix}
        1.3\cdot10^{-3}&0&0\\
        2.4\cdot10^{-5}&1.1\cdot10^{-5}&0\\
        1.8\cdot10^{-3}&1.1\cdot10^{-4}&1.0\cdot10^{-3}
    \end{pmatrix}
    \,,
    \\
\text{IO:}&\hspace{0.5cm}
    \abs{\mathbb{I}-\alpha}<\begin{pmatrix}
        1.4\cdot10^{-3}&0&0\\
        2.4\cdot10^{-5}&1.0\cdot10^{-5}&0\\
        1.6\cdot10^{-3}&3.6\cdot10^{-5}&8.1\cdot10^{-4}
    \end{pmatrix}
    \,.
\end{align}

\section{Global fit bounds for the general neutrino case (G-SS)}
\label{Sec:ResultsGSS}

When the seesaw scenario features more than three neutrinos, the dim-5 and 6 operators are in general independent (see {\it e.g.}~Refs.~\cite{Casas:2001sr,Broncano:2002rw,Antusch:2009gn,Blennow:2011vn}) and therefore there are no correlations in the flavor structure of the mixing coming from the correct reproduction of neutrino masses and mixings. The Schwarz inequality will in general not be saturated either, in contrast to the previous scenarios. Thus, the 6 elements of the $\eta$-matrix are completely independent and cLFV observables cannot be used to constrain the LFC ones. Therefore, we can extract the bounds on the diagonal elements $\eta_{\alpha\alpha}$ via a global fit to the observables of Table~\ref{table:observables} and, separately, extract the bounds on the off-diagonal entries from the bounds on the cLFV observables using Eq.~\eqref{eq:LFVrad}.
Additionally, the Schwarz inequality given by Eq.~\eqref{eq:schwarz} also allows us to derive constraints on the off-diagonal elements from the bounds on the diagonal ones. 

\begin{figure}[t!]
    \centering
    \includegraphics[width=0.95\textwidth]{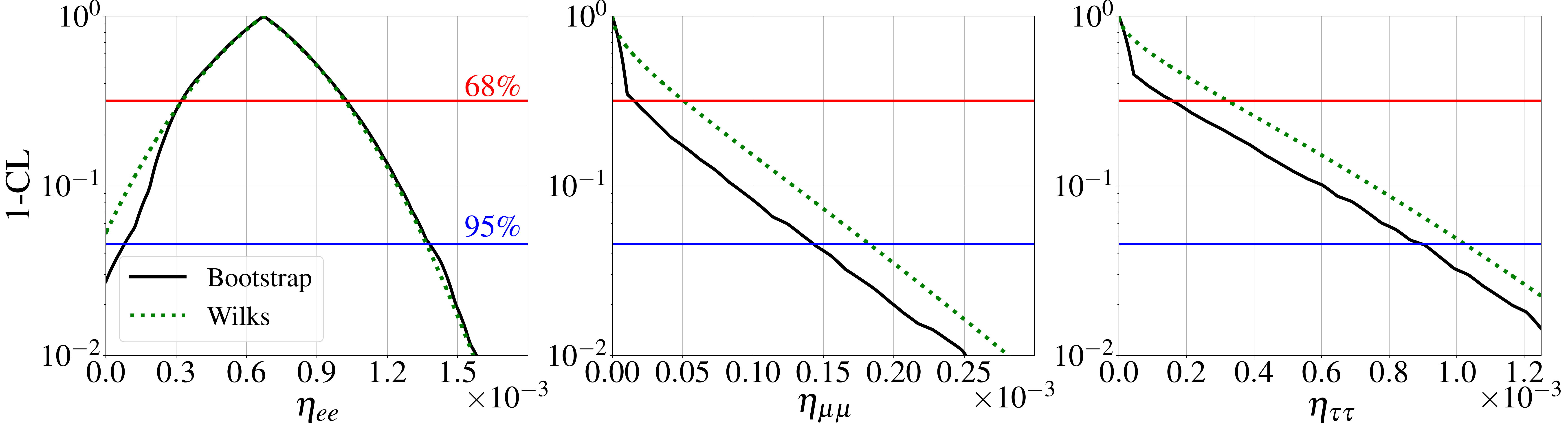}
    \caption{
    Same as Fig.~\ref{fig:2NSSprofiles} but for the general neutrino scenario with an arbitrary number of RH neutrinos (G-SS).}
    \label{fig:GSSprofiles}
\end{figure}

The results of our global fit are shown in Fig.~\ref{fig:GSSprofiles} and Table~\ref{table:GSSbounds}.
As in previous sections, Fig.~\ref{fig:GSSprofiles} shows the profiles obtained following Wilks' theorem and those obtained after calibrating the test statistic. 
Note that, despite the absence of correlations induced by reproducing the observed light neutrino mass and mixing, the $\eta_{\alpha\alpha}\geq0$ condition still holds, as required by Eq.~\eqref{eq:eta_theta}, which imposes a physical boundary and violates the requirements for Wilks' theorem to apply.
Therefore, we could expect that calibration effects will be more important close to the $\eta_{\alpha\alpha}=0$ border. Indeed, this can be clearly appreciated in the $\etaee$ profile, where the most substantial deviations appear near the physical boundary at zero. 
As we move away from $\etaee=0$, the two profiles converge, as naively expected. 
However, when moving away from the boundary, the $\etamumu$ and $\etatautau$ profiles do not seem to converge to Wilks'.
This is because $\etamumu$ and $\etatautau$ have a best-fit at zero and, thus, comparatively larger values would be required in order to recover the Wilks' behaviour.

The profiles show a slight preference ($\approx 2\sigma$) for a non-zero value of $\etaee$, whereas $\etamumu$ has quite stringent bounds. This is again driven by the Cabibbo anomaly in the CKM sector tightly constraining the size of $\etamumu$ and several observables pushing for a non-zero $\etaee$. Additionally, $\etatautau$ is less tightly constrained than $\etamumu$, mainly because the $\tau$ sector is constrained by less observables and does not contribute to the CKM anomaly. The preference for a non-zero $\etaee$ induces also a preference for a non-zero $\Tr{\eta}$.

Regarding the off-diagonal elements, we show their corresponding profiles in App.~\ref{App:Procedure}, and collect their bounds in Table~\ref{table:GSSbounds}. The second and third columns collect the bounds from the global fit to LFC observables including the Schwarz inequality. In the fourth and fith columns we show the constraints on the off-diagonal elements directly imposed by cLFV processes.
We have highlighted in bold the strongest bound for each sector. 
As can be seen, the cLFV bound is the most constraining in the $e$-$\mu$ sector due to the stringent upper limits on $\mu\to e$ transitions. 
Conversely, in the $e$-$\tau$ and $\mu$-$\tau$ sectors, the indirect LFC bounds from the Schwarz inequality clearly dominate.
Notice the very strong $68\%$CL bound on $\abs{\etamutau}$. This is a consequence of the strong border effects that appear since both $\etamumu$ and $\etatautau$ have their best-fit at $0$. As expected, the $95\%$CL bound is, in turn, much less tight.

\begin{table}
\centering
    \begin{tabular}{|c|c|c|c|c|}
    \hline 
    &\multicolumn{2}{c|}{}&\multicolumn{2}{c|}{}\\[-2.5ex]
    \multirow{2}{*}{\bf G-SS}&\multicolumn{2}{c|}{LFC Bound} 
    &\multicolumn{2}{c|}{LFV Bound}\\[0ex]  
    \cline{2-5} &&&&\\[-2.5ex]
    & $68\%$CL& $95\%$CL
    & $68\%$CL& $95\%$CL \\
    \hline
    &&&&\\[-2ex]
    $\eta_{ee}$ &
    $[0.33, 1.0] \cdot10^{-3}$ & $[0.081, 1.4] \cdot10^{-3}$ & -&- \\[1.2ex]
    $\eta_{\mu\mu}$ & $1.5 \cdot10^{-5}$ & $1.4 \cdot10^{-4}$ & - &-\\[1.2ex]
    $\eta_{\tau\tau}$  & $1.6 \cdot10^{-4}$ & $8.9 \cdot10^{-4}$ &-&-\\[1.2ex]
    $\Tr{\eta}$ & $[0.28,1.2]\cdot10^{-3}$ & $2.1\cdot10^{-3}$ &-&-\\[1.2ex]
    $\abs{\eta_{e\mu}}$ & $1.4\cdot10^{-4}$ & $3.4\cdot10^{-4}$ & $\mathbf{8.4\cdot10^{-6}}$ & $\mathbf{1.2\cdot10^{-5}}$ \\[1.2ex]
    $\abs{\eta_{e\tau}}$ & $\mathbf{4.2\cdot10^{-4}}$ & $\mathbf{8.8\cdot10^{-4}}$ & $5.7\cdot10^{-3}$ & $8.1\cdot10^{-3}$ \\[1.2ex]
    $\abs{\eta_{\mu\tau}}$ & $\mathbf{9.4\cdot10^{-6}}$ & $\mathbf{1.8\cdot10^{-4}}$ & $6.6\cdot10^{-3}$ & $9.4\cdot10^{-3}$ \\[1.2ex]
    
    \hline    
    \end{tabular}
    \caption{Upper bounds (or preferred intervals) for the G-SS.
    The LFC bounds are obtained from the global fit analysis to the observables in Table~\ref{table:observables}, in particular from the boostrapped profiles in Fig.~\ref{fig:GNUprofiles} (see also Fig.~\ref{fig:GSSprofiles_comparison} in the appendix).
    For off-diagonal $\eta_{\alpha\beta}$ elements, we also derive limits from cLFV transitions and highlight the strongest bound for each flavor sector.
    }
    \label{table:GSSbounds}
\end{table}

As before, we can translate the bounds to the $\alpha$ parametrization, which at $95\%$CL reads
\begin{equation}
    \abs{\mathbb{I}-\alpha}=\begin{pmatrix}
        [0.081, 1.4] \cdot10^{-3}&0&0\\
        <2.4\cdot10^{-5}&<1.4\cdot10^{-4}&0\\
        <1.8\cdot10^{-3}&<3.6\cdot10^{-4}&<8.9\cdot10^{-4}
    \end{pmatrix}.
\end{equation}

Finally, also in this more general setup the $\eta$ parametrization can be connected to the mixing between active and RH neutrinos, as given in Eq.~\eqref{eq:eta_theta}.
Consequently, the bounds in Table~\ref{table:GSSbounds} can be understood as bounds on the total mixings to a given flavor, defined as the sum of (squared) mixings of each RH neutrinos to a given flavor:
\begin{equation}
    \eta_{\alpha\alpha} = \frac12 |\Theta_\alpha|^2\,,
    \qquad
    {\rm with}
    \quad
    |\Theta_\alpha|^2 \equiv \sum_{k=1}^n |\Theta_{\alpha k}|^2\,.
\end{equation}
Thus, the bounds on these total mixings $\Theta_\alpha$ can be compared to those from other experimental facilities directly searching for heavy neutrinos (see {\it e.g.}~Ref.~\cite{Fernandez-Martinez:2023phj,Antel:2023hkf}), bearing in mind that they depend on the RH neutrino mass scale and apply to lighter scales than the ones considered here. 
The only caveat is that most of these experimental bounds for a given flavor are obtained switching-off the mixing with the other flavors, while we marginalized over them. Nevertheless, we have verified than considering 
the simplified scenario in which there is mixing exclusively with
one given flavor leads to very similar results to those in Table~\ref{table:GSSbounds}.

\section{Global fit bounds for generic unitarity violation (GUV)}
\label{Sec:ResultsGNU}

Finally, we consider the case in which the $\eta$-matrix is not assumed to be positive-definite. Thus, its diagonal entries are allowed to be also negative while
the off-diagonal parameters are not required to satisfy the Schwarz's inequality. 
This generic unitarity violation (GUV) cannot be realised in a pure Type-I Seesaw mechanism and a more elaborate particle content, with different contributions to non-unitarity would be required~\cite{Coutinho:2019aiy}.
This extra freedom can lead to rather different results from the G-SS setup due to the Cabibbo anomaly being solved by negative values of $\etamumu$, which are unattainable for a Type-I Seesaw but are possible for an unconstrained $\eta$-matrix~\cite{Coutinho:2019aiy}. 

From the point of view of the analysis, the LFC and LFV fits can be decoupled as the diagonal and off-diagonal elements are not correlated, similarly to the G-SS. However, contrary to the G-SS case, the off-diagonal entries can only be constrained from LFV bounds, as the Schwarz's inequality no longer holds. Furthermore, in the GUV setup there are no physical boundaries and the $\eta_{\alpha\alpha}$ can cover both positive and negative values. This means that Wilks' theorem can be safely assumed, as the violations present in the G-SS are no longer an issue.

It should be noted that, contrary to the previous cases, in this scenario we have adopted an agnostic stance on the kind of new physics responsible for non-unitarity. As such, the extra degrees of freedom that may contribute to the loop level cLFV observables are unknown. We thus consider exclusively the contribution of the light neutrinos to the cLFV processes in order to obtain a bound on $\abs{\eta_{\alpha\beta}}$, which assumes that no substantial cancellations with the contributions from possible new particles.  In particular, the contribution of only the light neutrinos to radiative decays is:
\begin{equation}\label{eq:LFVrad_lightneutrinos}
{\rm BR}(\ell_\alpha\to \ell_\beta\gamma)= \frac{25\alpha}{6\pi}\,\big|\eta_{\alpha\beta}\big|^2\,,
\end{equation}
as opposed to Eq.~\eqref{eq:LFVrad} which contains also the heavy neutrino contributions. Since we are neglecting the contributions from the new particles, the off-diagonal bounds we will quote using Eq.~\eqref{eq:LFVrad_lightneutrinos} are only orientative and model-dependent.

We perform our statistical analysis for the LFC observables leaving the sign of $\eta_{\alpha\alpha}$ unconstrained. Remarkably, a much better fit to the data than in the G-SS is found. Indeed, while the G-SS improves the fit with respect to the SM by $\Delta\chi^2=3.75$, the GUV setup improves upon the SM fit by $\Delta\chi^2=11.07$. As previously stated, this is mainly due to the fact that the GUV scenario can fit the Cabibbo anomaly, whereas the G-SS cannot.
\begin{figure}[t!]
    \centering
    \includegraphics[width=0.95\textwidth]{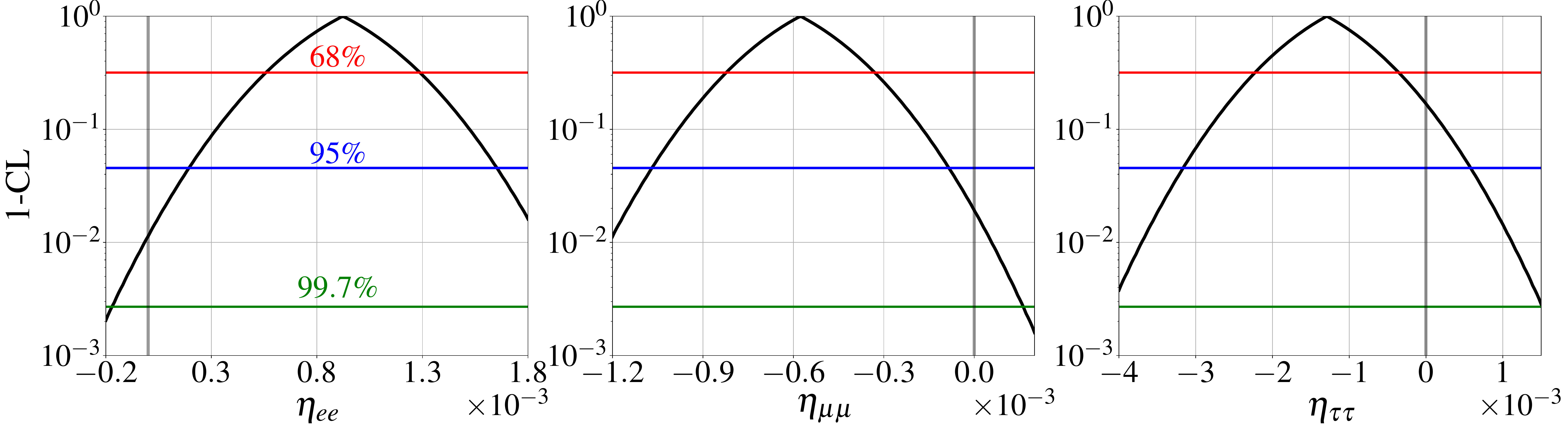}
    \caption{Same as Fig.~\ref{fig:2NSSprofiles} but for the generic unitarity violation scenario (GUV), where the $\eta$-matrix can also take negative values and Eq.~\eqref{eq:schwarz} does not need to be satisfied.}
    \label{fig:GNUprofiles}
\end{figure}

The resulting profiles are shown in Fig.~\ref{fig:GNUprofiles}, where bootstrapping provides no deviation from Wilks' theorem, as expected.
Apart from the $2\sigma$ preference for a non-zero $\etaee$, which was already present in the G-SS, we also find a $2\sigma$ preference for $\etamumu < 0$ induced by the Cabibbo anomaly. 
Conversely, we do not find any significance preference for $\etatautau$ different from zero. The corresponding preferred intervals and upper bounds are collected in Table~\ref{table:GNUbounds}.

Finally, as in the previous cases, these bounds can be translated to the $\alpha$ parametrization. At $95\%$CL, we have:
\begin{align}
    1-\alpha_{11}& \in [0.20,1.65]\cdot10^{-3}\,, & \abs{\alpha_{21}}&<1.4\cdot10^{-5}\,,\nonumber
    \\
    1-\alpha_{22}& \in [-1.1,-0.088]\cdot10^{-3}\,, & \abs{\alpha_{31}}&<9.8\cdot10^{-3}\,,
    \\
    1-\alpha_{33}& \in [-3.1,0.56]\cdot10^{-3}\,, & \abs{\alpha_{32}}&<1.1\cdot10^{-2}\,.\nonumber
\end{align}

\begin{table}[t!]
\centering
    \begin{tabular}{|c|c|c||c|c|c|}
    \hline 
    &\multicolumn{2}{c||}{}&&\multicolumn{2}{c|}{}\\[-2.5ex]
    \multirow{2}{*}{\bf GUV}&\multicolumn{2}{c||}{LFC Bound} 
    &\multirow{2}{*}{}&\multicolumn{2}{c|}{LFV Bound}\\[0ex]  
    \cline{2-3}
    \cline{5-6} &&&&&\\[-2.5ex]
    & $68\%$CL& $95\%$CL&
    & $68\%$CL& $95\%$CL \\
    \hline
    &&&&&\\[-2ex]
    $\eta_{ee}$ & $[0.56, 1.29] \cdot10^{-3}$ & $[0.20, 1.65] \cdot10^{-3}$ & 
    $\abs{\eta_{e\mu}}$  & $5.0\cdot10^{-6}$ & $7.2\cdot10^{-6}$ \\[1.2ex]
    $\eta_{\mu\mu}$ & $[-8.2,-3.3]\cdot10^{-4}$ & $[-1.1,-0.088] \cdot10^{-3}$ &
    $\abs{\eta_{e\tau}}$  & $3.4\cdot10^{-3}$ & $4.9\cdot10^{-3}$ \\[1.2ex]
    $\eta_{\tau\tau}$  & $[-2.2,-0.38] \cdot10^{-3}$ & $[-3.1,0.56] \cdot10^{-3}$ &
    $\abs{\eta_{\mu\tau}}$ & $4.0\cdot10^{-3}$ & $5.6\cdot10^{-3}$ \\[1.2ex]
    \hline    
    \end{tabular}
    \caption{Upper bounds (or preferred intervals) for the GUV.}
    \label{table:GNUbounds}
\end{table}

\section{Discussion and Conclusions}
\label{Sec:Discussion}

In this work we have updated and improved upon present constraints on the unitarity of the leptonic mixing matrix and the mixing of heavy right-handed neutrinos with the SM active flavours with a global fit to flavour and electroweak precision observables. Besides updating all the experimental constraints from the different observables under consideration, compared to previous studies we have improved the analysis in several additional ways. 
In particular, given the expected deviations from Wilks' theorem, we have explicitly calibrated our test statistics by boostrapping in order to properly estimate the significance of the constraints placed. We have also explicitly computed the SM prediction for the different observables in terms of the input data instead of using the results of the available electroweak fits, as has been done in the past, so as to make the global fit fully consistent.

We have provided results for four distinct scenarios. The first three correspond to leptonic mixing unitarity deviations induced by scenarios with 2, 3 or a higher arbitrary number of extra right-handed neutrinos, dubbed 2N-SS, 3N-SS, and G-SS, respectively. 
The results are summarized in Tables~\ref{table:2NSSbounds}, \ref{table:3NSSbounds}, and \ref{table:GSSbounds}.
The main difference among these scenarios is the level of correlation between the parameters that describe the unitarity deviations, provided that the correct pattern of neutrino masses and mixings is also recovered. The deviations from unitarity are encoded in all generality through a small Hermitian matrix $\eta$. This matrix corresponds to the coefficient of the only $d=6$ operator obtained at tree level upon integrating out the heavy neutrinos, and is moreover directly connected with the square of their mixing with the active neutrino flavours. Additionally, we also considered the possibility of lepton mixing unitarity deviations not necessarily induced by the presence of additional right-handed neutrinos, whose results are collected in Table~\ref{table:GNUbounds}. This scenario removes the requirement that they are described by a positive definite matrix $\eta$. For all the scenarios considered, we also report our results in an alternative parametrization of unitarity deviations through a lower triangular matrix, more appropriate when studying the neutrino oscillation phenomenon.

We derived our results from a global fit to the observables summarized in Table~\ref{table:observables}. Thus, the set of bounds derived is valid as long as the new degrees of freedom are heavier than the mass of the $Z$. For lighter new particles, some of the LEP constraints at the $Z$ pole are lost, but the remaining observables apply down to the mass of the $\tau$. Notice that we did not include the latest CDF-II measurement on $M_W$ among our observables since it is in significant tension (above $5\sigma$) with the rest of measurements with the same parametric dependence on $\eta$ (see section~\ref{sec:DCF}).

The strongest constraint on unitarity deviations comes from the very stringent bound from $\mu \to e$ transitions. In the 2N-SS scenario this observable dominates the constraints on essentially all elements, given the little freedom and strong correlations implied by having only 2 right-handed neutrinos reproducing the measured pattern of light neutrino masses and mixings. In the 3N-SS, $\mu \to e$ processes instead forbids mixing to both electrons and muons simultaneously, while in the more general scenarios it provides strong constraints on the $\etaemu$ element. Conversely, constraints from $\tau \to e$ and $\tau \to \mu$ transitions are generally subleading and stronger bounds are implied through the LFC observables. Nevertheless, when unitarity deviations are not sourced by right-handed neutrinos exclusively, both sets of constraints would, a priori, be uncorrelated cLFV observables are the only bounds applying on $\etaetau$ and $\etamutau$.

Regarding the lepton flavor conserving observables we find the following behaviours: Due to the existing tension in the unitarity test of the first row of the CKM (the so-called Cabibbo anomaly), $\eta_{\mu \mu} > 0$ (or equivalently right-handed neutrino mixing with muons $\theta_\mu \neq 0$, if that is the source of $\eta$) is disfavoured, since its presence worsens the anomaly. Moreover, even though LFU observables are generally in good agreement with the SM, there is a mild preference ($\sim 1\sigma$) for $\etaee > \etamumu$, both in $\LFUratio{\mu}{e}{\pi}$ and $\LFUratio{\mu}{e}{\tau}.$ On the other hand, the measured values of $s^2_{\text{eff}}$ and $M_W$ show a slight preference ($\sim1$-$2\sigma$) for a deviation from unitarity either in the electron or muon sector.

The combination of these effects leads to constraints on all the $\eta$ parameters ranging between $10^{-3}$ and $10^{-5}$ at $2\sigma$ for both the 3N-SS and G-SS, with the exception of a preference for non-unitarity at the level of $\eta_{ee}\sim 10^{-3}$ at around $2\sigma$ (see Figs.~\ref{fig:3NSSprofile} and \ref{fig:GSSprofiles}). This implies a $2\sigma$ preference for a mixing of the heavy neutrinos with the electron at the $|\theta_{e}|\sim 10^{-2}$ level. Conversely, unitarity deviations are very disfavoured in the case of the muons due to the Cabibbo anomaly and the LFU constraints. Finally, the tau sector is almost exclusively constrained by the Z-pole observables $\Gamma_{\text{inv}}$, $\Gamma_Z$ and $\sigma^{0}_{\text{had}}$ which show no preference for non-unitarity.

The main difference regarding the 2N-SS is that stronger bounds on the $\eta$ elements, ranging approximately between $10^{-4}$ and $10^{-5}$ at $2\sigma$, are found given its more constrained structure. Indeed, for normal neutrino mass ordering there is not enough freedom to fit the preference for a dominant mixing with the electron flavor. Thus, slightly stronger constraints and no preference for unitarity violations are found with respect to the inverted ordering case. Conversely, for inverted ordering (or when more than 2 right-handed neutrinos are considered), the constraints implied by the measured pattern of light neutrino masses and mixings are compatible with a dominant role of the mixing to the electron flavor, as data prefers. Thus, these scenarios provide a somewhat better fit with a preference for unitarity violation in the electron sector.

Finally, if the unitarity deviations are not assumed to be positive definite (as required when they are solely induced by right-handed neutrinos), negative values in the muon sector may actually solve the Cabibbo anomaly. Thus, in this most general, but also more complex, extension a significantly better fit is found with a preference for unitarity deviations ($\eta_{\alpha\alpha}\neq 0$) at the $2 \sigma$ level both in the electron and muon sectors and with opposite signs.

\medskip

\paragraph{Acknowledgments.}
 The authors thank Gonzalo Morrás for very illuminating discussions.  
This project has received support from the European Union’s Horizon 2020 research and innovation programme under the Marie Skłodowska-Curie grant agreement No~860881-HIDDeN and No 101086085 - ASYMMETRY, and from the Spanish Research Agency (Agencia Estatal de Investigaci\'on) through the Grant IFT Centro de Excelencia Severo Ochoa No CEX2020-001007-S and Grant PID2019-108892RB-I00 funded by MCIN/AEI/10.13039/501100011033. EFM, XM and DNT acknowledge support from the HPC-Hydra cluster at IFT.
XM acknowledges funding from the European Union’s Horizon Europe Programme under the Marie Skłodowska-Curie grant agreement no.~101066105-PheNUmenal. The work of DNT was supported by the
Spanish MIU through the National Program FPU (grant number FPU20/05333). JLP also acknowledges support from Generalitat Valenciana through the plan GenT program (CIDEGENT/2018/019) and from the Spanish Ministerio de Ciencia e Innovacion through the project PID2020-113644GB-I00. JHG warmly thanks the hospitality of Albert De Roeck and the EP Neutrino group during his stay at CERN; where this project has been completed.

\appendix

\section{Details of the analysis}
\label{App:Procedure}
Wilks' theorem is a common assumption in many statistical analyses. It establishes that the test statistic on the form
\begin{equation}
    \Delta \chi^2 = \chi^2(\theta) - \min_{\theta'} \chi^2(\theta')\,,
\end{equation}
where
\begin{equation}
    \chi^2(\theta) = \sum_{d \in D} \left(\frac{\mu_d(\theta) - \bar\mu_d}{\sigma_d}\right)^2,
\end{equation}
$\theta$ is a set of model parameters, $D$ the set of data points, $\mu_d(\theta)$ the expected data for $d$ given $\theta$, $\bar\mu_d$ the observed data, and $\sigma_d^2$ the variance in $\mu_d(\theta)$, is everywhere distributed according to a $\chi^2$-distribution, with a number of degrees of freedom that can be derived from the total number of free parameters. When the theorem applies, it can be used to directly map the test statistic to a confidence level.

However, for the theorem to hold, several assumptions on the behaviour of the test statistic have to be made. One of the most important assumptions, which is generally violated in our analysis, is that the expectation $\mu_d(\theta)$ for the observables included in the fit must describe a hyperplane in the space of possible observations as one varies the model parameters $\theta$. A straight-forward source of violation of Wilks' theorem assumptions is that our parameters have a physical boundary ($\eta_{\alpha\alpha}\geq 0$) in the 2N-SS, 3N-SS and G-SS scenarios, and therefore so will our observables. Moreover, we will also find violations when including LFV observables, since they depend quadratically on $\eta$ unlike to the LFC ones of Table~\ref{table:observables}, which depend linearly on $\eta$, thereby leading to a curved expected region of the observations. 
Additionally, when the observables depend on cyclic parameters, such as phases, they will describe compact trajectories in the space of possible observations, providing and additional source of violation of Wilks' theorem.

Our analysis is therefore plagued with potential deviations from Wilks' theorem, and we cannot rely on it for extracting confidence intervals. Instead, the test statistic must be calibrated at each point of parameter space, in order to know how it is distributed and associate the resulting value of $\Delta\chi^2$ to a confidence level. This procedure is commonly referred to as \textit{bootstrapping}. We describe the bootstrapping procedure of our analysis below.

We start by computing the $\chi^2(\eta_{\alpha \beta})$ where $\eta_{\alpha \beta}$ are the parameters of the model assuming Gaussian uncertainties for all the experimental values of the observables $O_i$ reported in Table~\ref{table:observables} and taking into account correlations among them when relevant:
\begin{equation}
\chi^2(\eta_{\alpha\beta}) = \sum_{i,j} \frac{\big(O_i- E_i(\eta_{\alpha \beta})\big)}{\sigma_i} \mathrm{corr}_{ij} \frac{\big(O_j- E_j(\eta_{\alpha \beta})\big)}{\sigma_j}\,,
\end{equation}
where $E_i(\eta_{\alpha \beta})$ is the expectation of $O_i$ given the model parameters $\eta_{\alpha \beta}$ as given by the first two columns in Table~\ref{table:observables}, $\sigma_i$ is the corresponding uncertainty reported together with $O_i$ in the third column of Table~\ref{table:observables} and $\mathrm{corr}_{ij}$ is the correlation matrix among the different observables.

Since we are interested on showing the CL profiles for a particular parameter of interest, for definiteness $\eta_{e e}$ but the procedure would be the same for any other parameter, we choose the profiled $\chi^2$ over all the parameters $\eta_{\alpha \beta}$ except for $\eta_{e e}$ itself as our test statistic. Denoting the other (nuisance) parameters $\vec{\nu} = \eta_{\alpha \beta}$ with $\alpha \beta \neq ee$, then the profiled $\chi^2$ is:
\begin{equation}
    \hat{\chi}^2(\eta_{e e})\equiv \min_{\vec{\nu}} \chi^2(\eta_{e e},\vec{\nu})\,. 
\end{equation}
If Wilks' theorem holds, then the significance with which a point $\eta_{e e}^0$ may be excluded by the observables $O_i$ will be given by the square root of:
\begin{equation}
  \Delta\chi^2(\eta_{e e}^0)= \hat{\chi}^2(\eta_{e e}^0)-\chi^2_{\text{min}}\,,
    \label{app:deltachi2}
\end{equation}
where $\chi^2_{\text{min}}$ is the minimum value of the test statistic, since $\hat{\chi}^2(\eta_{e e})$ would follow a $\chi^2$ distribution with 1 degree of freedom. Note that only the value of $\eta_{e e}^0$ is relevant independently of in which point of nuisance parameter space $\vec{\nu}$ the statement is made, as Wilks' theorem guarantees that the test statistic of Eq.~\eqref{app:deltachi2} is $\chi^2$-distributed (with one degree of freedom) in all the parameter space. However, in the presence of violations of Wilks' theorem, there is no reason why this should be the case and we cannot assume a $\chi^2$-distribution for the test statistic. Moreover, the distribution of the test statistic may be different in different points of parameter space.

Instead, in order to claim an exclusion confidence level for $\eta_{e e}^0$, we need to consider all points in the $\vec{\nu}$ parameter space, calibrate the test statistic's distribution in each of these points, and then compare each of these calibrated distributions with the actual value $\Delta\chi^2$ from Eq.~\eqref{app:deltachi2} in order to extract a confidence level. The minimum of these confidence levels would then represent the significance of the exclusion of $\eta_{e e}^0$ regardless of $\vec \eta$.

We therefore implement the following bootstrap procedure:
\begin{enumerate}
    \item The values of the parameter of interest for which we want to quote a confidence level are fixed, $\{\eta_{e e}^i\}_{i=1}^n$. For each of those values, we consider a grid of points in $\vec{\nu}$ space. 
    \item For each of the $\eta_{e e}^i$, we consider the different $\vec\nu$ and compute the corresponding predictions for the observables of the fit.
    \item Using the predictions as the central value, we generate pseudo-data for the observables by drawing from a Gaussian distribution.
    \item For each of the sampled sets of pseudo-data, we compute the corresponding value of the test statistic $\Delta\chi^2(\eta_{e e}^i)$.
\end{enumerate}
Repeating the generation of pseudo-data and consequent computation of the test statistic many times results in a large sample of the test statistic distribution. A confidence level is then computed by comparing the real data with this distribution, the CL simply being the percentage of times that $\Delta\chi^2(\eta_{e e}^i)$ of the pseudo-data sets are smaller than the value computed for the real data. Repeating this procedure for all $\vec{\nu}$ points for a certain $\eta_{e e}^i$ value and then picking the minimum CL, we can finally associate an exclusion level to $\eta_{e e}^i$.

Since we need to generate a large amount of pseudo-data over a grid of points in parameter space, the number of points required grows exponentially with the number of parameters. Moreover, since at each point of the grid many numerical minimizations of the test statistic are required in order to reconstruct its distribution, the computational cost of this procedure quickly becomes unfeasible. Nevertheless, for the G-SS and 2N-SS scenarios, this procedure is still computationally tractable, since their parameter spaces are three-dimensional. However, in the case of the 3N-SS, characterized by 8 free parameters, this is no longer the case.

We have therefore opted for an approximation to the bootstrap procedure for the 3N-SS scenario, which we dub \textit{Profiled Bootstrap}. As it name suggests, it consists on only performing the calibration in the points that profile the test statistic along the direction of some parameter of interest $\eta_{e e}$. As such, for each value $\eta_{e e}^i$, instead of considering a grid of points in nuisance parameter space $\vec{\nu}$, we will only consider the point:
\begin{equation}
    \hat{\vec{\nu}}(\eta_{e e}^i)\equiv \underset{\vec{\nu}}{\text{argmin }} \chi^2(\eta_{e e}^i,\vec{\nu})\,.
\end{equation}
We are thus reducing the task of exploring a grid of points in $\vec{\nu}$ space to just a single point. The rest of the procedure follows steps 2-4 as before.

\begin{figure}[t!]
    \centering
    \includegraphics[width=\textwidth]{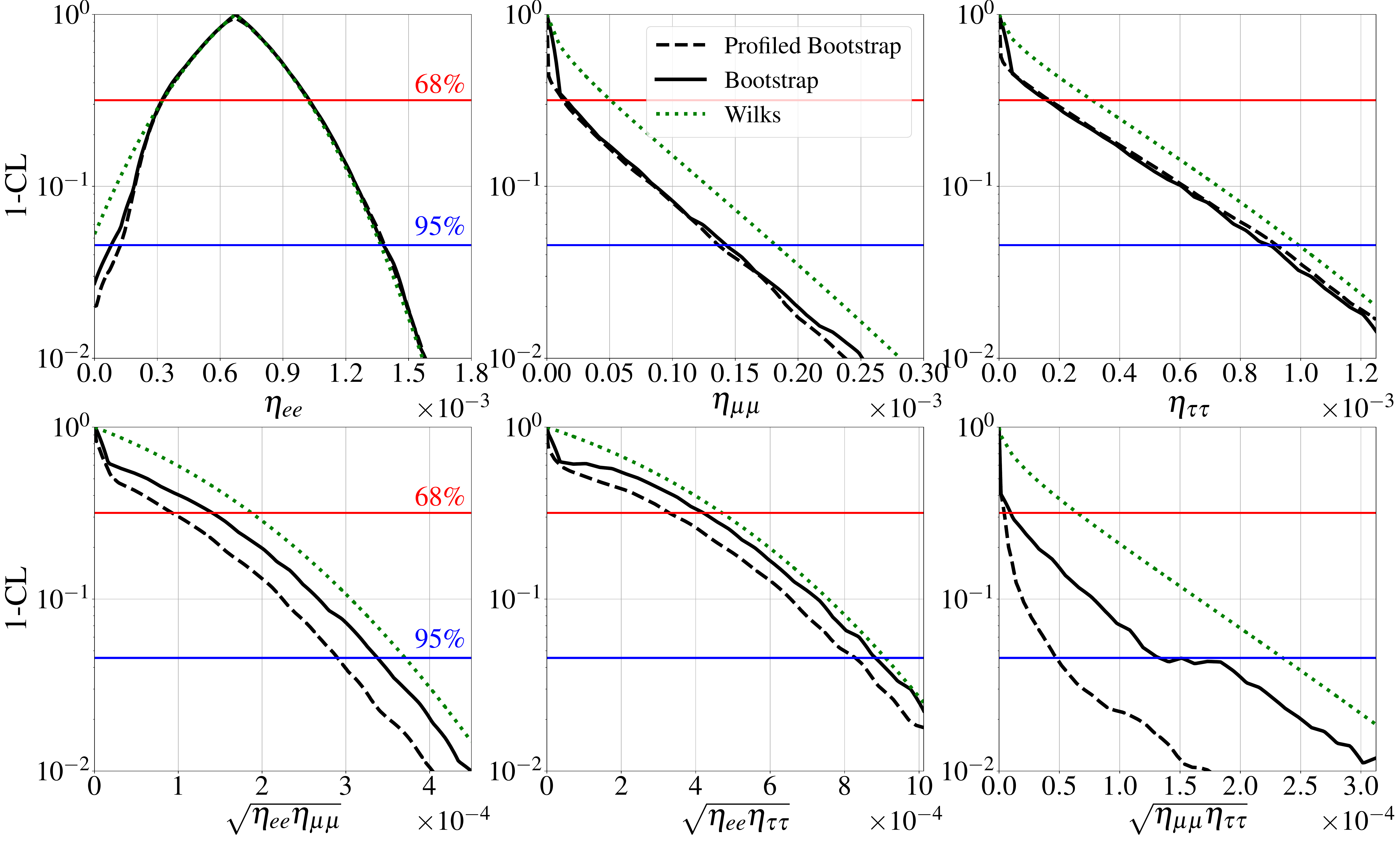}
    \caption{Comparison between the bootstrap (solid black) and profiled bootstrap (dashed black) for the G-SS case. 
    We also show the expectations of Wilks' theorem (dashed green) for reference. In general we see quite good agreement between both procedures for the diagonal elements in the upper panels, but not for the off-diagonal combinations in the lower panels.}
    \label{fig:GSSprofiles_comparison}
\end{figure}

The requirement that needs to be satisfied in order for the profiled bootstrap to be an acceptable approximation to the complete bootstrap is that the point $\hat{\vec{\nu}}(\eta_{e e}^i)$ needs to be close to the point in $\vec{\nu}$ space that yields the minimum CL for a fixed $\eta_{e e}^i$. While this can seem to intuitively always be the case, there may be scenarios in which points in the parameter space with a value of the test statistic far from its minimum may yield a smaller CL if the distribution of the test statistic leans toward even larger values in that point of the parameter space. We have thus compared the original bootstrap and the profiled bootstrap in the G-SS case, which is a scenario in which both procedures are tractable, to test its accuracy. The results are shown in the upper panels of Fig.~\ref{fig:GSSprofiles_comparison} where we can see that, in general, both the solid and dashed black lines are in good agreement.
The reason why this behaviour is expected to also hold for the 3N-SS is that  both the G-SS and 3N-SS showcase very similar freedom when it comes to fitting the data: even though the 3N-SS case presents non trivial correlations between the $\eta_{\alpha\alpha}$ parameters (see Eq.~\eqref{3NSS:mixing_correlation}), they do not strongly constrain the flavor pattern\footnote{This is not the case of the 2N-SS, which showcases a quite predictive flavor pattern. However, it should be noted the flavor pattern of the 3N-SS becomes more predictive for $\mlightest$ values much smaller than the existing bounds~\cite{Chrzaszcz:2019inj}.}, as it can be seen in Fig.~\ref{fig:triangles}.

However, while the agreement between these two procedures is good for the profiles of the diagonal elements $\eta_{\alpha\alpha}$, this is no longer the case for the off-diagonal combinations $\sqrt{\eta_{\alpha\alpha}\eta_{\beta\beta}}$, as can be seen in the lower panels of Fig.~\ref{fig:GSSprofiles_comparison}. Therefore, assuming the same discrepancy will be present for the 3N-SS, we adopted a more conservative approach and the off-diagonal bounds quoted in Table~\ref{table:3NSSbounds} are extracted assuming Wilks' theorem, which provides the weaker constraint between the two options. Note that the latter only applies for the 3N-SS, since in the 2N-SS and G-SS scenarios we are able to follow the full bootstrap procedure also for the off-diagonal elements.

\section{About cancellations in the cLFV rates}
\label{App:Cancelations}

In section~\ref{Sec:cLFV} we neglected potential differences between the masses of all heavy neutrinos, so we could simplify the discussion of cLFV transitions. 
The bounds we derived after including these rates should still apply to more generic scenarios, nevertheless in principle having several mass scales could lead to potential cancellations in some of the cLFV rates, as advocated in Ref.~\cite{Forero:2011pc}. We devote this appendix to such a discussion.

First of all, let us emphasize that this discussion does not apply to the minimal scenarios in sections~\ref{Sec:Results2N} and \ref{Sec:Results3N}, where the Schwarz inequality in Eq.~\eqref{eq:schwarz} is saturated, meaning that the cLFV bounds must be included in our global fit and that they impose bounds also for the diagonal entries of the $\eta$-matrix.   
The reason is that there is only a single relevant heavy scale and, moreover, the flavor mixing pattern is quite defined by oscillation data. 
Thus, there is not enough freedom to accommodate the potential cancellations we discuss here, and the bounds derived for those scenarios are robust. 

The question about potential cancellations in the cLFV rates becomes relevant for models with more heavy neutrinos, for which the general bounds in section~\ref{Sec:ResultsGSS} should be respected, and when having several heavy neutrino mass scales.
In such a scenario, and taking the radiative decays as an example, we have
\begin{equation}
    \Gamma(\ell_\alpha\to\ell_\beta \gamma)\propto \left|\sum_{k} \Theta_{\alpha k} \Theta^*_{\beta k}\, F_\gamma\Big(M_{N_k}^2/M_W^2\Big)\right|^2\,,
\end{equation}
where the sum goes over all the heavy neutrinos, and the missing factors and explicit form of the loop-function $F_\gamma$ can be found for instance in Ref.~\cite{Alonso:2012ji}.

If all heavy neutrinos are degenerate or, more generally, if the value of $F_\gamma\Big(M_{N_k}^2/M_W^2\Big)$ is common for all contributions, the sum of $\Theta_{\alpha k} \Theta^*_{\beta k}$ gives the $\eta_{\alpha\beta}$ dependence in Eq.~\eqref{eq:eta_theta}. Otherwise, additional dependence on other parameters beyond $\eta_{\alpha \beta}$ will be present, potentially even leading to a suppression of the radiative decay for non-vanishing $\eta_{\alpha\beta}$. 
Nevertheless, the loop function monotonically increases from $1/8$ to $1/2$ as the heavy neutrino mass varies from $M_W$ to infinity. Thus, deviations from a constant value for $F_\gamma\Big(M_{N_k}^2/M_W^2\Big)$ cannot be large. Furthermore, a cancellation requires opposite-sign contributions from the mixing terms $\Theta^{}_{\alpha k} \Theta^*_{\beta k}$, which would also (partially) suppress $\eta_{\alpha\beta}$.
Therefore, having a large $\abs{\eta_{\alpha\beta}}$ with suppressed $\ell_\alpha\to\ell_\beta \gamma$ is possible only for rather large $\Theta^{}_{\alpha k} \Theta^*_{\beta k}$ mixings and for very specific values of the heavy neutrino masses, so that is only a partial cancellation in $\abs{\eta_{\alpha\beta}}$ but the values of $F_\gamma$ are such that the radiative decays are strongly suppressed.

Furthermore, this kind of configurations would lead to a numerical cancellation only for the radiative decays, but in general they will not suppress other cLFV observables such as the 3-body decays or $\mu-e$ conversion in nuclei, which can actually be more restrictive, as we have seen in section~\ref{Sec:cLFV}. 
While adding more neutrinos enhances the number of parameters and the freedom to choose very specific values of mixings and masses so that all the cLFV channels are suppressed (but not the respective $\abs{\eta_{\alpha\beta}}$), this seems an extremely unlikely configuration.

Nevertheless, to account also for this possibility, our results for the G-SS in section \ref{Sec:ResultsGSS} are given considering separately the lepton flavor conserving and violating channels, see Table~\ref{table:GSSbounds}.
While for $\abs{\eta_{e\mu}}$ the bounds from the cLFV observables are stronger, they may be regarded as more model dependent, as discussed above. Conversely, LFC bounds, which are also the most stringent for $\abs{\eta_{e\tau}}$ and $\abs{\eta_{\mu\tau}}$ cannot be avoided by these kind of cancellations.


\bibliographystyle{JHEP} 
\bibliography{biblio}

\providecommand{\href}[2]{#2}\begingroup\raggedright\begin{thebibliography}{100}

\bibitem{Dodelson:1993je}
S.~Dodelson and L.~M. Widrow, {\it {Sterile-neutrinos as dark matter}},  {\em
  Phys. Rev. Lett.} {\bf 72} (1994) 17--20,
  [\href{http://arxiv.org/abs/hep-ph/9303287}{{\tt hep-ph/9303287}}].

\bibitem{Shi:1998km}
X.-D. Shi and G.~M. Fuller, {\it {A New dark matter candidate: Nonthermal
  sterile neutrinos}},  {\em Phys. Rev. Lett.} {\bf 82} (1999) 2832--2835,
  [\href{http://arxiv.org/abs/astro-ph/9810076}{{\tt astro-ph/9810076}}].

\bibitem{Abazajian:2001nj}
K.~Abazajian, G.~M. Fuller, and M.~Patel, {\it {Sterile neutrino hot, warm, and
  cold dark matter}},  {\em Phys. Rev. D} {\bf 64} (2001) 023501,
  [\href{http://arxiv.org/abs/astro-ph/0101524}{{\tt astro-ph/0101524}}].

\bibitem{Asaka:2005an}
T.~Asaka, S.~Blanchet, and M.~Shaposhnikov, {\it {The nuMSM, dark matter and
  neutrino masses}},  {\em Phys. Lett. B} {\bf 631} (2005) 151--156,
  [\href{http://arxiv.org/abs/hep-ph/0503065}{{\tt hep-ph/0503065}}].

\bibitem{Falkowski:2009yz}
A.~Falkowski, J.~Juknevich, and J.~Shelton, {\it {Dark Matter Through the
  Neutrino Portal}},  \href{http://arxiv.org/abs/0908.1790}{{\tt
  arXiv:0908.1790}}.

\bibitem{Lindner:2010rr}
M.~Lindner, A.~Merle, and V.~Niro, {\it {Enhancing Dark Matter Annihilation
  into Neutrinos}},  {\em Phys. Rev. D} {\bf 82} (2010) 123529,
  [\href{http://arxiv.org/abs/1005.3116}{{\tt arXiv:1005.3116}}].

\bibitem{GonzalezMacias:2015rxl}
V.~Gonzalez~Macias and J.~Wudka, {\it {Effective theories for Dark Matter
  interactions and the neutrino portal paradigm}},  {\em JHEP} {\bf 07} (2015)
  161, [\href{http://arxiv.org/abs/1506.03825}{{\tt arXiv:1506.03825}}].

\bibitem{Blennow:2019fhy}
M.~Blennow, E.~Fernandez-Martinez, A.~Olivares-Del~Campo, S.~Pascoli,
  S.~Rosauro-Alcaraz, and A.~V. Titov, {\it {Neutrino Portals to Dark Matter}},
   {\em Eur. Phys. J. C} {\bf 79} (2019), no.~7 555,
  [\href{http://arxiv.org/abs/1903.00006}{{\tt arXiv:1903.00006}}].

\bibitem{Fukugita:1986hr}
M.~Fukugita and T.~Yanagida, {\it {Baryogenesis Without Grand Unification}},
  {\em Phys. Lett. B} {\bf 174} (1986) 45--47.

\bibitem{Akhmedov:1998qx}
E.~K. Akhmedov, V.~A. Rubakov, and A.~Y. Smirnov, {\it {Baryogenesis via
  neutrino oscillations}},  {\em Phys. Rev. Lett.} {\bf 81} (1998) 1359--1362,
  [\href{http://arxiv.org/abs/hep-ph/9803255}{{\tt hep-ph/9803255}}].

\bibitem{Asaka:2005pn}
T.~Asaka and M.~Shaposhnikov, {\it {The $\nu$MSM, dark matter and baryon
  asymmetry of the universe}},  {\em Phys. Lett. B} {\bf 620} (2005) 17--26,
  [\href{http://arxiv.org/abs/hep-ph/0505013}{{\tt hep-ph/0505013}}].

\bibitem{Drewes:2017zyw}
M.~Drewes, B.~Garbrecht, P.~Hernandez, M.~Kekic, J.~Lopez-Pavon, J.~Racker,
  N.~Rius, J.~Salvado, and D.~Teresi, {\it {ARS Leptogenesis}},  {\em Int. J.
  Mod. Phys. A} {\bf 33} (2018), no.~05n06 1842002,
  [\href{http://arxiv.org/abs/1711.02862}{{\tt arXiv:1711.02862}}].

\bibitem{Antel:2023hkf}
C.~Antel et~al., {\it {Feebly Interacting Particles: FIPs 2022 workshop
  report}},  in {\em {Workshop on Feebly-Interacting Particles}}, 5, 2023.
\newblock \href{http://arxiv.org/abs/2305.01715}{{\tt arXiv:2305.01715}}.

\bibitem{Fernandez-Martinez:2023phj}
E.~Fern\'andez-Mart\'\i{}nez, M.~Gonz\'alez-L\'opez,
  J.~Hern\'andez-Garc\'\i{}a, M.~Hostert, and J.~L\'opez-Pav\'on, {\it
  {Effective portals to heavy neutral leptons}},
  \href{http://arxiv.org/abs/2304.06772}{{\tt arXiv:2304.06772}}.

\bibitem{MatheusRepository}
M.~Hostert. {\it Heavy Neutrino Limits} GitHub repository,
  \url{https://github.com/mhostert/Heavy-Neutrino-Limits}.

\bibitem{Petcov:1976ff}
S.~T. Petcov, {\it {The Processes $\mu \rightarrow e + \gamma, \mu \rightarrow
  e + \overline{e}, \nu' \rightarrow \nu + \gamma$ in the Weinberg-Salam Model
  with Neutrino Mixing}},  {\em Sov. J. Nucl. Phys.} {\bf 25} (1977) 340.
  [Erratum: Sov.J.Nucl.Phys. 25, 698 (1977), Erratum: Yad.Fiz. 25, 1336
  (1977)].

\bibitem{Bilenky:1977du}
S.~M. Bilenky, S.~T. Petcov, and B.~Pontecorvo, {\it {Lepton Mixing, mu
  --\ensuremath{>} e + gamma Decay and Neutrino Oscillations}},  {\em Phys.
  Lett. B} {\bf 67} (1977) 309.

\bibitem{Cheng:1977vn}
T.-P. Cheng and L.-F. Li, {\it {Muon Number Nonconservation in Gauge
  Theories}},  {\em Stud. Nat. Sci.} {\bf 12} (1977) 659--681.

\bibitem{Marciano:1977wx}
W.~J. Marciano and A.~I. Sanda, {\it {Exotic Decays of the Muon and Heavy
  Leptons in Gauge Theories}},  {\em Phys. Lett. B} {\bf 67} (1977) 303--305.

\bibitem{Lee:1977qz}
B.~W. Lee, S.~Pakvasa, R.~E. Shrock, and H.~Sugawara, {\it {Muon and Electron
  Number Nonconservation in a V-A Gauge Model}},  {\em Phys. Rev. Lett.} {\bf
  38} (1977) 937. [Erratum: Phys.Rev.Lett. 38, 1230 (1977)].

\bibitem{Lee:1977tib}
B.~W. Lee and R.~E. Shrock, {\it {Natural Suppression of Symmetry Violation in
  Gauge Theories: Muon - Lepton and Electron Lepton Number Nonconservation}},
  {\em Phys. Rev. D} {\bf 16} (1977) 1444.

\bibitem{Shrock:1980vy}
R.~E. Shrock, {\it {New Tests For, and Bounds On, Neutrino Masses and Lepton
  Mixing}},  {\em Phys. Lett. B} {\bf 96} (1980) 159--164.

\bibitem{Schechter:1980gr}
J.~Schechter and J.~W.~F. Valle, {\it {Neutrino Masses in SU(2) x U(1)
  Theories}},  {\em Phys. Rev. D} {\bf 22} (1980) 2227.

\bibitem{Shrock:1980ct}
R.~E. Shrock, {\it {General Theory of Weak Leptonic and Semileptonic Decays. 1.
  Leptonic Pseudoscalar Meson Decays, with Associated Tests For, and Bounds on,
  Neutrino Masses and Lepton Mixing}},  {\em Phys. Rev. D} {\bf 24} (1981)
  1232.

\bibitem{Shrock:1981wq}
R.~E. Shrock, {\it {General Theory of Weak Processes Involving Neutrinos. 2.
  Pure Leptonic Decays}},  {\em Phys. Rev. D} {\bf 24} (1981) 1275.

\bibitem{Langacker:1988ur}
P.~Langacker and D.~London, {\it {Mixing Between Ordinary and Exotic
  Fermions}},  {\em Phys. Rev. D} {\bf 38} (1988) 886.

\bibitem{Pilaftsis:1992st}
A.~Pilaftsis, {\it {Lepton flavor nonconservation in H0 decays}},  {\em Phys.
  Lett. B} {\bf 285} (1992) 68--74.

\bibitem{Ilakovac:1994kj}
A.~Ilakovac and A.~Pilaftsis, {\it {Flavor violating charged lepton decays in
  seesaw-type models}},  {\em Nucl. Phys. B} {\bf 437} (1995) 491,
  [\href{http://arxiv.org/abs/hep-ph/9403398}{{\tt hep-ph/9403398}}].

\bibitem{Nardi:1994iv}
E.~Nardi, E.~Roulet, and D.~Tommasini, {\it {Limits on neutrino mixing with new
  heavy particles}},  {\em Phys. Lett. B} {\bf 327} (1994) 319--326,
  [\href{http://arxiv.org/abs/hep-ph/9402224}{{\tt hep-ph/9402224}}].

\bibitem{Tommasini:1995ii}
D.~Tommasini, G.~Barenboim, J.~Bernabeu, and C.~Jarlskog, {\it {Nondecoupling
  of heavy neutrinos and lepton flavor violation}},  {\em Nucl. Phys. B} {\bf
  444} (1995) 451--467, [\href{http://arxiv.org/abs/hep-ph/9503228}{{\tt
  hep-ph/9503228}}].

\bibitem{Illana:2000ic}
J.~I. Illana and T.~Riemann, {\it {Charged lepton flavor violation from massive
  neutrinos in Z decays}},  {\em Phys. Rev. D} {\bf 63} (2001) 053004,
  [\href{http://arxiv.org/abs/hep-ph/0010193}{{\tt hep-ph/0010193}}].

\bibitem{Loinaz:2003gc}
W.~Loinaz, N.~Okamura, S.~Rayyan, T.~Takeuchi, and L.~C.~R. Wijewardhana, {\it
  {Quark lepton unification and lepton flavor nonconservation from a TeV scale
  seesaw neutrino mass texture}},  {\em Phys. Rev. D} {\bf 68} (2003) 073001,
  [\href{http://arxiv.org/abs/hep-ph/0304004}{{\tt hep-ph/0304004}}].

\bibitem{Arganda:2004bz}
E.~Arganda, A.~M. Curiel, M.~J. Herrero, and D.~Temes, {\it {Lepton flavor
  violating Higgs boson decays from massive seesaw neutrinos}},  {\em Phys.
  Rev. D} {\bf 71} (2005) 035011,
  [\href{http://arxiv.org/abs/hep-ph/0407302}{{\tt hep-ph/0407302}}].

\bibitem{Loinaz:2004qc}
W.~Loinaz, N.~Okamura, S.~Rayyan, T.~Takeuchi, and L.~C.~R. Wijewardhana, {\it
  {The NuTeV anomaly, lepton universality, and nonuniversal neutrino gauge
  couplings}},  {\em Phys. Rev. D} {\bf 70} (2004) 113004,
  [\href{http://arxiv.org/abs/hep-ph/0403306}{{\tt hep-ph/0403306}}].

\bibitem{Antusch:2006vwa}
S.~Antusch, C.~Biggio, E.~Fernandez-Martinez, M.~B. Gavela, and J.~Lopez-Pavon,
  {\it {Unitarity of the Leptonic Mixing Matrix}},  {\em JHEP} {\bf 10} (2006)
  084, [\href{http://arxiv.org/abs/hep-ph/0607020}{{\tt hep-ph/0607020}}].

\bibitem{Antusch:2008tz}
S.~Antusch, J.~P. Baumann, and E.~Fernandez-Martinez, {\it {Non-Standard
  Neutrino Interactions with Matter from Physics Beyond the Standard Model}},
  {\em Nucl. Phys. B} {\bf 810} (2009) 369--388,
  [\href{http://arxiv.org/abs/0807.1003}{{\tt arXiv:0807.1003}}].

\bibitem{Biggio:2008in}
C.~Biggio, {\it {The Contribution of fermionic seesaws to the anomalous
  magnetic moment of leptons}},  {\em Phys. Lett. B} {\bf 668} (2008) 378--384,
  [\href{http://arxiv.org/abs/0806.2558}{{\tt arXiv:0806.2558}}].

\bibitem{Alonso:2012ji}
R.~Alonso, M.~Dhen, M.~B. Gavela, and T.~Hambye, {\it {Muon conversion to
  electron in nuclei in type-I seesaw models}},  {\em JHEP} {\bf 01} (2013)
  118, [\href{http://arxiv.org/abs/1209.2679}{{\tt arXiv:1209.2679}}].

\bibitem{Abada:2012mc}
A.~Abada, D.~Das, A.~M. Teixeira, A.~Vicente, and C.~Weiland, {\it {Tree-level
  lepton universality violation in the presence of sterile neutrinos: impact
  for $R_K$ and $R_\pi$}},  {\em JHEP} {\bf 02} (2013) 048,
  [\href{http://arxiv.org/abs/1211.3052}{{\tt arXiv:1211.3052}}].

\bibitem{Akhmedov:2013hec}
E.~Akhmedov, A.~Kartavtsev, M.~Lindner, L.~Michaels, and J.~Smirnov, {\it
  {Improving Electro-Weak Fits with TeV-scale Sterile Neutrinos}},  {\em JHEP}
  {\bf 05} (2013) 081, [\href{http://arxiv.org/abs/1302.1872}{{\tt
  arXiv:1302.1872}}].

\bibitem{Basso:2013jka}
L.~Basso, O.~Fischer, and J.~J. van~der Bij, {\it {Precision tests of unitarity
  in leptonic mixing}},  {\em EPL} {\bf 105} (2014), no.~1 11001,
  [\href{http://arxiv.org/abs/1310.2057}{{\tt arXiv:1310.2057}}].

\bibitem{Abada:2013aba}
A.~Abada, A.~M. Teixeira, A.~Vicente, and C.~Weiland, {\it {Sterile neutrinos
  in leptonic and semileptonic decays}},  {\em JHEP} {\bf 02} (2014) 091,
  [\href{http://arxiv.org/abs/1311.2830}{{\tt arXiv:1311.2830}}].

\bibitem{Arganda:2014dta}
E.~Arganda, M.~J. Herrero, X.~Marcano, and C.~Weiland, {\it {Imprints of
  massive inverse seesaw model neutrinos in lepton flavor violating Higgs boson
  decays}},  {\em Phys. Rev. D} {\bf 91} (2015), no.~1 015001,
  [\href{http://arxiv.org/abs/1405.4300}{{\tt arXiv:1405.4300}}].

\bibitem{Antusch:2014woa}
S.~Antusch and O.~Fischer, {\it {Non-unitarity of the leptonic mixing matrix:
  Present bounds and future sensitivities}},  {\em JHEP} {\bf 10} (2014) 094,
  [\href{http://arxiv.org/abs/1407.6607}{{\tt arXiv:1407.6607}}].

\bibitem{Antusch:2015mia}
S.~Antusch and O.~Fischer, {\it {Testing sterile neutrino extensions of the
  Standard Model at future lepton colliders}},  {\em JHEP} {\bf 05} (2015) 053,
  [\href{http://arxiv.org/abs/1502.05915}{{\tt arXiv:1502.05915}}].

\bibitem{Abada:2014cca}
A.~Abada, V.~De~Romeri, S.~Monteil, J.~Orloff, and A.~M. Teixeira, {\it
  {Indirect searches for sterile neutrinos at a high-luminosity Z-factory}},
  {\em JHEP} {\bf 04} (2015) 051, [\href{http://arxiv.org/abs/1412.6322}{{\tt
  arXiv:1412.6322}}].

\bibitem{Abada:2015oba}
A.~Abada, V.~De~Romeri, and A.~M. Teixeira, {\it {Impact of sterile neutrinos
  on nuclear-assisted cLFV processes}},  {\em JHEP} {\bf 02} (2016) 083,
  [\href{http://arxiv.org/abs/1510.06657}{{\tt arXiv:1510.06657}}].

\bibitem{Abada:2015trh}
A.~Abada and T.~Toma, {\it {Electric Dipole Moments of Charged Leptons with
  Sterile Fermions}},  {\em JHEP} {\bf 02} (2016) 174,
  [\href{http://arxiv.org/abs/1511.03265}{{\tt arXiv:1511.03265}}].

\bibitem{Abada:2016awd}
A.~Abada and T.~Toma, {\it {Electron electric dipole moment in Inverse Seesaw
  models}},  {\em JHEP} {\bf 08} (2016) 079,
  [\href{http://arxiv.org/abs/1605.07643}{{\tt arXiv:1605.07643}}].

\bibitem{DeRomeri:2016gum}
V.~De~Romeri, M.~J. Herrero, X.~Marcano, and F.~Scarcella, {\it {Lepton flavor
  violating Z decays: A promising window to low scale seesaw neutrinos}},  {\em
  Phys. Rev. D} {\bf 95} (2017), no.~7 075028,
  [\href{http://arxiv.org/abs/1607.05257}{{\tt arXiv:1607.05257}}].

\bibitem{Arganda:2016zvc}
E.~Arganda, M.~J. Herrero, X.~Marcano, R.~Morales, and A.~Szynkman, {\it
  {Effective lepton flavor violating H\ensuremath{\ell}i\ensuremath{\ell}j
  vertex from right-handed neutrinos within the mass insertion approximation}},
   {\em Phys. Rev. D} {\bf 95} (2017), no.~9 095029,
  [\href{http://arxiv.org/abs/1612.09290}{{\tt arXiv:1612.09290}}].

\bibitem{Herrero:2018luu}
M.~J. Herrero, X.~Marcano, R.~Morales, and A.~Szynkman, {\it {One-loop
  effective LFV $Zl_kl_m$ vertex from heavy neutrinos within the mass insertion
  approximation}},  {\em Eur. Phys. J. C} {\bf 78} (2018), no.~10 815,
  [\href{http://arxiv.org/abs/1807.01698}{{\tt arXiv:1807.01698}}].

\bibitem{Marcano:2019rmk}
X.~Marcano and R.~A. Morales, {\it {Flavor techniques for LFV processes: Higgs
  decays in a general seesaw model}},  {\em Front. in Phys.} {\bf 7} (2020)
  228, [\href{http://arxiv.org/abs/1909.05888}{{\tt arXiv:1909.05888}}].

\bibitem{Chrzaszcz:2019inj}
M.~Chrzaszcz, M.~Drewes, T.~E. Gonzalo, J.~Harz, S.~Krishnamurthy, and
  C.~Weniger, {\it {A frequentist analysis of three right-handed neutrinos with
  GAMBIT}},  {\em Eur. Phys. J. C} {\bf 80} (2020), no.~6 569,
  [\href{http://arxiv.org/abs/1908.02302}{{\tt arXiv:1908.02302}}].

\bibitem{Coutinho:2019aiy}
A.~M. Coutinho, A.~Crivellin, and C.~A. Manzari, {\it {Global Fit to Modified
  Neutrino Couplings and the Cabibbo-Angle Anomaly}},  {\em Phys. Rev. Lett.}
  {\bf 125} (2020), no.~7 071802, [\href{http://arxiv.org/abs/1912.08823}{{\tt
  arXiv:1912.08823}}].

\bibitem{Calderon:2022alb}
K.~A.~U. Calder\'on, I.~Timiryasov, and O.~Ruchayskiy, {\it {Improved
  constraints and the prospects of detecting TeV to PeV scale Heavy Neutral
  Leptons}},  \href{http://arxiv.org/abs/2206.04540}{{\tt arXiv:2206.04540}}.

\bibitem{Weinberg:1979sa}
S.~Weinberg, {\it {Baryon and Lepton Nonconserving Processes}},  {\em Phys.
  Rev. Lett.} {\bf 43} (1979) 1566--1570.

\bibitem{Broncano:2002rw}
A.~Broncano, M.~B. Gavela, and E.~E. Jenkins, {\it {The Effective Lagrangian
  for the seesaw model of neutrino mass and leptogenesis}},  {\em Phys. Lett.
  B} {\bf 552} (2003) 177--184,
  [\href{http://arxiv.org/abs/hep-ph/0210271}{{\tt hep-ph/0210271}}]. [Erratum:
  Phys.Lett.B 636, 332 (2006)].

\bibitem{Minkowski:1977sc}
P.~Minkowski, {\it {$\mu \to e\gamma$ at a Rate of One Out of $10^{9}$ Muon
  Decays?}},  {\em Phys. Lett. B} {\bf 67} (1977) 421--428.

\bibitem{Mohapatra:1979ia}
R.~N. Mohapatra and G.~Senjanovic, {\it {Neutrino Mass and Spontaneous Parity
  Nonconservation}},  {\em Phys. Rev. Lett.} {\bf 44} (1980) 912.

\bibitem{Yanagida:1979as}
T.~Yanagida, {\it {Horizontal gauge symmetry and masses of neutrinos}},  {\em
  Conf. Proc. C} {\bf 7902131} (1979) 95--99.

\bibitem{Gell-Mann:1979vob}
M.~Gell-Mann, P.~Ramond, and R.~Slansky, {\it {Complex Spinors and Unified
  Theories}},  {\em Conf. Proc. C} {\bf 790927} (1979) 315--321,
  [\href{http://arxiv.org/abs/1306.4669}{{\tt arXiv:1306.4669}}].

\bibitem{Branco:1988ex}
G.~C. Branco, W.~Grimus, and L.~Lavoura, {\it {The Seesaw Mechanism in the
  Presence of a Conserved Lepton Number}},  {\em Nucl. Phys. B} {\bf 312}
  (1989) 492--508.

\bibitem{Kersten:2007vk}
J.~Kersten and A.~Y. Smirnov, {\it {Right-Handed Neutrinos at CERN LHC and the
  Mechanism of Neutrino Mass Generation}},  {\em Phys. Rev. D} {\bf 76} (2007)
  073005, [\href{http://arxiv.org/abs/0705.3221}{{\tt arXiv:0705.3221}}].

\bibitem{Abada:2007ux}
A.~Abada, C.~Biggio, F.~Bonnet, M.~B. Gavela, and T.~Hambye, {\it {Low energy
  effects of neutrino masses}},  {\em JHEP} {\bf 12} (2007) 061,
  [\href{http://arxiv.org/abs/0707.4058}{{\tt arXiv:0707.4058}}].

\bibitem{Moffat:2017feq}
K.~Moffat, S.~Pascoli, and C.~Weiland, {\it {Equivalence between massless
  neutrinos and lepton number conservation in fermionic singlet extensions of
  the Standard Model}},  \href{http://arxiv.org/abs/1712.07611}{{\tt
  arXiv:1712.07611}}.

\bibitem{Mohapatra:1986aw}
R.~N. Mohapatra, {\it {Mechanism for Understanding Small Neutrino Mass in
  Superstring Theories}},  {\em Phys. Rev. Lett.} {\bf 56} (1986) 561--563.

\bibitem{Mohapatra:1986bd}
R.~N. Mohapatra and J.~W.~F. Valle, {\it {Neutrino Mass and Baryon Number
  Nonconservation in Superstring Models}},  {\em Phys. Rev. D} {\bf 34} (1986)
  1642.

\bibitem{Akhmedov:1995ip}
E.~K. Akhmedov, M.~Lindner, E.~Schnapka, and J.~W.~F. Valle, {\it {Left-right
  symmetry breaking in NJL approach}},  {\em Phys. Lett. B} {\bf 368} (1996)
  270--280, [\href{http://arxiv.org/abs/hep-ph/9507275}{{\tt hep-ph/9507275}}].

\bibitem{Malinsky:2005bi}
M.~Malinsky, J.~C. Romao, and J.~W.~F. Valle, {\it {Novel supersymmetric SO(10)
  seesaw mechanism}},  {\em Phys. Rev. Lett.} {\bf 95} (2005) 161801,
  [\href{http://arxiv.org/abs/hep-ph/0506296}{{\tt hep-ph/0506296}}].

\bibitem{Fernandez-Martinez:2007iaa}
E.~Fernandez-Martinez, M.~B. Gavela, J.~Lopez-Pavon, and O.~Yasuda, {\it
  {CP-violation from non-unitary leptonic mixing}},  {\em Phys. Lett. B} {\bf
  649} (2007) 427--435, [\href{http://arxiv.org/abs/hep-ph/0703098}{{\tt
  hep-ph/0703098}}].

\bibitem{Gavela:2009cd}
M.~B. Gavela, T.~Hambye, D.~Hernandez, and P.~Hernandez, {\it {Minimal Flavour
  Seesaw Models}},  {\em JHEP} {\bf 09} (2009) 038,
  [\href{http://arxiv.org/abs/0906.1461}{{\tt arXiv:0906.1461}}].

\bibitem{Fernandez-Martinez:2015hxa}
E.~Fernandez-Martinez, J.~Hernandez-Garcia, J.~Lopez-Pavon, and M.~Lucente,
  {\it {Loop level constraints on Seesaw neutrino mixing}},  {\em JHEP} {\bf
  10} (2015) 130, [\href{http://arxiv.org/abs/1508.03051}{{\tt
  arXiv:1508.03051}}].

\bibitem{Fernandez-Martinez:2016lgt}
E.~Fernandez-Martinez, J.~Hernandez-Garcia, and J.~Lopez-Pavon, {\it {Global
  constraints on heavy neutrino mixing}},  {\em JHEP} {\bf 08} (2016) 033,
  [\href{http://arxiv.org/abs/1605.08774}{{\tt arXiv:1605.08774}}].

\bibitem{CDF:2022hxs}
{\bf CDF} Collaboration, T.~Aaltonen et~al., {\it {High-precision measurement
  of the $W$ boson mass with the CDF II detector}},  {\em Science} {\bf 376}
  (2022), no.~6589 170--176.

\bibitem{ParticleDataGroup:2022pth}
{\bf Particle Data Group} Collaboration, R.~L. Workman et~al., {\it {Review of
  Particle Physics}},  {\em PTEP} {\bf 2022} (2022) 083C01.

\bibitem{CMS:2022ett}
{\bf CMS} Collaboration, {\it {Precision measurement of the Z boson invisible
  width in pp collisions at $\sqrt{s}$ = 13 TeV}},
  \href{http://arxiv.org/abs/2206.07110}{{\tt arXiv:2206.07110}}.

\bibitem{Bryman:2021teu}
D.~Bryman, V.~Cirigliano, A.~Crivellin, and G.~Inguglia, {\it {Testing Lepton
  Flavor Universality with Pion, Kaon, Tau, and Beta Decays}},
  \href{http://arxiv.org/abs/2111.05338}{{\tt arXiv:2111.05338}}.

\bibitem{HFLAV:2019otj}
{\bf HFLAV} Collaboration, Y.~S. Amhis et~al., {\it {Averages of b-hadron,
  c-hadron, and $\tau $-lepton properties as of 2018}},  {\em Eur. Phys. J. C}
  {\bf 81} (2021), no.~3 226, [\href{http://arxiv.org/abs/1909.12524}{{\tt
  arXiv:1909.12524}}].

\bibitem{Awramik:2003rn}
M.~Awramik, M.~Czakon, A.~Freitas, and G.~Weiglein, {\it {Precise prediction
  for the W boson mass in the standard model}},  {\em Phys. Rev. D} {\bf 69}
  (2004) 053006, [\href{http://arxiv.org/abs/hep-ph/0311148}{{\tt
  hep-ph/0311148}}].

\bibitem{Awramik:2006uz}
M.~Awramik, M.~Czakon, and A.~Freitas, {\it {Electroweak two-loop corrections
  to the effective weak mixing angle}},  {\em JHEP} {\bf 11} (2006) 048,
  [\href{http://arxiv.org/abs/hep-ph/0608099}{{\tt hep-ph/0608099}}].

\bibitem{ATLAS:2023fsi}
{\bf ATLAS} Collaboration, {\it {Improved W boson Mass Measurement using 7 TeV
  Proton-Proton Collisions with the ATLAS Detector}}, .

\bibitem{LHCb:2021bjt}
{\bf LHCb} Collaboration, R.~Aaij et~al., {\it {Measurement of the W boson
  mass}},  {\em JHEP} {\bf 01} (2022) 036,
  [\href{http://arxiv.org/abs/2109.01113}{{\tt arXiv:2109.01113}}].

\bibitem{CDF:2013dpa}
{\bf CDF, D0} Collaboration, T.~A. Aaltonen et~al., {\it {Combination of CDF
  and D0 $W$-Boson Mass Measurements}},  {\em Phys. Rev. D} {\bf 88} (2013),
  no.~5 052018, [\href{http://arxiv.org/abs/1307.7627}{{\tt arXiv:1307.7627}}].

\bibitem{ALEPH:2013dgf}
{\bf ALEPH, DELPHI, L3, OPAL, LEP Electroweak} Collaboration, S.~Schael et~al.,
  {\it {Electroweak Measurements in Electron-Positron Collisions at
  W-Boson-Pair Energies at LEP}},  {\em Phys. Rept.} {\bf 532} (2013) 119--244,
  [\href{http://arxiv.org/abs/1302.3415}{{\tt arXiv:1302.3415}}].

\bibitem{LYONS1988110}
L.~Lyons, D.~Gibaut, and P.~Clifford, {\it How to combine correlated estimates
  of a single physical quantity},  {\em Nuclear Instruments and Methods in
  Physics Research Section A: Accelerators, Spectrometers, Detectors and
  Associated Equipment} {\bf 270} (1988), no.~1 110--117.

\bibitem{VALASSI2003391}
A.~Valassi, {\it Combining correlated measurements of several different
  physical quantities},  {\em Nuclear Instruments and Methods in Physics
  Research Section A: Accelerators, Spectrometers, Detectors and Associated
  Equipment} {\bf 500} (2003), no.~1 391--405. NIMA Vol 500.

\bibitem{Janot:2019oyi}
P.~Janot and S.~Jadach, {\it {Improved Bhabha cross section at LEP and the
  number of light neutrino species}},  {\em Phys. Lett. B} {\bf 803} (2020)
  135319, [\href{http://arxiv.org/abs/1912.02067}{{\tt arXiv:1912.02067}}].

\bibitem{Freitas:2014hra}
A.~Freitas, {\it {Higher-order electroweak corrections to the partial widths
  and branching ratios of the Z boson}},  {\em JHEP} {\bf 04} (2014) 070,
  [\href{http://arxiv.org/abs/1401.2447}{{\tt arXiv:1401.2447}}].

\bibitem{Baikov:2012er}
P.~A. Baikov, K.~G. Chetyrkin, J.~H. Kuhn, and J.~Rittinger, {\it {Complete
  ${\cal O}(\alpha_s^4)$ QCD Corrections to Hadronic $Z$-Decays}},  {\em Phys.
  Rev. Lett.} {\bf 108} (2012) 222003,
  [\href{http://arxiv.org/abs/1201.5804}{{\tt arXiv:1201.5804}}].

\bibitem{Seng:2018yzq}
C.-Y. Seng, M.~Gorchtein, H.~H. Patel, and M.~J. Ramsey-Musolf, {\it {Reduced
  Hadronic Uncertainty in the Determination of $V_{ud}$}},  {\em Phys. Rev.
  Lett.} {\bf 121} (2018), no.~24 241804,
  [\href{http://arxiv.org/abs/1807.10197}{{\tt arXiv:1807.10197}}].

\bibitem{Czarnecki:2019mwq}
A.~Czarnecki, W.~J. Marciano, and A.~Sirlin, {\it {Radiative Corrections to
  Neutron and Nuclear Beta Decays Revisited}},  {\em Phys. Rev. D} {\bf 100}
  (2019), no.~7 073008, [\href{http://arxiv.org/abs/1907.06737}{{\tt
  arXiv:1907.06737}}].

\bibitem{Seng:2020wjq}
C.-Y. Seng, X.~Feng, M.~Gorchtein, and L.-C. Jin, {\it {Joint lattice
  QCD\textendash{}dispersion theory analysis confirms the quark-mixing top-row
  unitarity deficit}},  {\em Phys. Rev. D} {\bf 101} (2020), no.~11 111301,
  [\href{http://arxiv.org/abs/2003.11264}{{\tt arXiv:2003.11264}}].

\bibitem{Grossman:2019bzp}
Y.~Grossman, E.~Passemar, and S.~Schacht, {\it {On the Statistical Treatment of
  the Cabibbo Angle Anomaly}},  {\em JHEP} {\bf 07} (2020) 068,
  [\href{http://arxiv.org/abs/1911.07821}{{\tt arXiv:1911.07821}}].

\bibitem{Blennow:2022yfm}
M.~Blennow, P.~Coloma, E.~Fern\'andez-Mart\'\i{}nez, and M.~Gonz\'alez-L\'opez,
  {\it {Right-handed neutrinos and the CDF II anomaly}},  {\em Phys. Rev. D}
  {\bf 106} (2022), no.~7 073005, [\href{http://arxiv.org/abs/2204.04559}{{\tt
  arXiv:2204.04559}}].

\bibitem{Maltoni:2003cu}
M.~Maltoni and T.~Schwetz, {\it {Testing the statistical compatibility of
  independent data sets}},  {\em Phys. Rev. D} {\bf 68} (2003) 033020,
  [\href{http://arxiv.org/abs/hep-ph/0304176}{{\tt hep-ph/0304176}}].

\bibitem{Glashow:1970gm}
S.~L. Glashow, J.~Iliopoulos, and L.~Maiani, {\it {Weak Interactions with
  Lepton-Hadron Symmetry}},  {\em Phys. Rev. D} {\bf 2} (1970) 1285--1292.

\bibitem{MEG:2016leq}
{\bf MEG} Collaboration, A.~M. Baldini et~al., {\it {Search for the lepton
  flavour violating decay $\mu ^+ \rightarrow \mathrm {e}^+ \gamma $ with the
  full dataset of the MEG experiment}},  {\em Eur. Phys. J. C} {\bf 76} (2016),
  no.~8 434, [\href{http://arxiv.org/abs/1605.05081}{{\tt arXiv:1605.05081}}].

\bibitem{BaBar:2009hkt}
{\bf BaBar} Collaboration, B.~Aubert et~al., {\it {Searches for Lepton Flavor
  Violation in the Decays tau+- ---\ensuremath{>} e+- gamma and tau+-
  ---\ensuremath{>} mu+- gamma}},  {\em Phys. Rev. Lett.} {\bf 104} (2010)
  021802, [\href{http://arxiv.org/abs/0908.2381}{{\tt arXiv:0908.2381}}].

\bibitem{Belle:2021ysv}
{\bf Belle} Collaboration, A.~Abdesselam et~al., {\it {Search for
  lepton-flavor-violating tau-lepton decays to $\ell\gamma$ at Belle}},  {\em
  JHEP} {\bf 10} (2021) 19, [\href{http://arxiv.org/abs/2103.12994}{{\tt
  arXiv:2103.12994}}].

\bibitem{SINDRUM:1987nra}
{\bf SINDRUM} Collaboration, U.~Bellgardt et~al., {\it {Search for the Decay
  mu+ ---\ensuremath{>} e+ e+ e-}},  {\em Nucl. Phys. B} {\bf 299} (1988) 1--6.

\bibitem{Hayasaka:2010np}
K.~Hayasaka et~al., {\it {Search for Lepton Flavor Violating Tau Decays into
  Three Leptons with 719 Million Produced Tau+Tau- Pairs}},  {\em Phys. Lett.
  B} {\bf 687} (2010) 139--143, [\href{http://arxiv.org/abs/1001.3221}{{\tt
  arXiv:1001.3221}}].

\bibitem{SINDRUMII:1993gxf}
{\bf SINDRUM II} Collaboration, C.~Dohmen et~al., {\it {Test of lepton flavor
  conservation in mu ---\ensuremath{>} e conversion on titanium}},  {\em Phys.
  Lett. B} {\bf 317} (1993) 631--636.

\bibitem{SINDRUMII:2006dvw}
{\bf SINDRUM II} Collaboration, W.~H. Bertl et~al., {\it {A Search for muon to
  electron conversion in muonic gold}},  {\em Eur. Phys. J. C} {\bf 47} (2006)
  337--346.

\bibitem{Forero:2011pc}
D.~V. Forero, S.~Morisi, M.~Tortola, and J.~W.~F. Valle, {\it {Lepton flavor
  violation and non-unitary lepton mixing in low-scale type-I seesaw}},  {\em
  JHEP} {\bf 09} (2011) 142, [\href{http://arxiv.org/abs/1107.6009}{{\tt
  arXiv:1107.6009}}].

\bibitem{Esteban:2020cvm}
I.~Esteban, M.~C. Gonzalez-Garcia, M.~Maltoni, T.~Schwetz, and A.~Zhou, {\it
  {The fate of hints: updated global analysis of three-flavor neutrino
  oscillations}},  {\em JHEP} {\bf 09} (2020) 178,
  [\href{http://arxiv.org/abs/2007.14792}{{\tt arXiv:2007.14792}}].

\bibitem{Caputo:2017pit}
A.~Caputo, P.~Hernandez, J.~Lopez-Pavon, and J.~Salvado, {\it {The seesaw
  portal in testable models of neutrino masses}},  {\em JHEP} {\bf 06} (2017)
  112, [\href{http://arxiv.org/abs/1704.08721}{{\tt arXiv:1704.08721}}].

\bibitem{Xing:2007zj}
Z.-z. Xing, {\it {Correlation between the Charged Current Interactions of Light
  and Heavy Majorana Neutrinos}},  {\em Phys. Lett. B} {\bf 660} (2008)
  515--521, [\href{http://arxiv.org/abs/0709.2220}{{\tt arXiv:0709.2220}}].

\bibitem{Escrihuela:2015wra}
F.~J. Escrihuela, D.~V. Forero, O.~G. Miranda, M.~Tortola, and J.~W.~F. Valle,
  {\it {On the description of nonunitary neutrino mixing}},  {\em Phys. Rev. D}
  {\bf 92} (2015), no.~5 053009, [\href{http://arxiv.org/abs/1503.08879}{{\tt
  arXiv:1503.08879}}]. [Erratum: Phys.Rev.D 93, 119905 (2016)].

\bibitem{Blennow:2016jkn}
M.~Blennow, P.~Coloma, E.~Fernandez-Martinez, J.~Hernandez-Garcia, and
  J.~Lopez-Pavon, {\it {Non-Unitarity, sterile neutrinos, and Non-Standard
  neutrino Interactions}},  {\em JHEP} {\bf 04} (2017) 153,
  [\href{http://arxiv.org/abs/1609.08637}{{\tt arXiv:1609.08637}}].

\bibitem{KATRIN:2021uub}
{\bf KATRIN} Collaboration, M.~Aker et~al., {\it {Direct neutrino-mass
  measurement with sub-electronvolt sensitivity}},  {\em Nature Phys.} {\bf 18}
  (2022), no.~2 160--166, [\href{http://arxiv.org/abs/2105.08533}{{\tt
  arXiv:2105.08533}}].

\bibitem{Planck:2018vyg}
{\bf Planck} Collaboration, N.~Aghanim et~al., {\it {Planck 2018 results. VI.
  Cosmological parameters}},  {\em Astron. Astrophys.} {\bf 641} (2020) A6,
  [\href{http://arxiv.org/abs/1807.06209}{{\tt arXiv:1807.06209}}]. [Erratum:
  Astron.Astrophys. 652, C4 (2021)].

\bibitem{Feldman:1997qc}
G.~J. Feldman and R.~D. Cousins, {\it {A Unified approach to the classical
  statistical analysis of small signals}},  {\em Phys. Rev. D} {\bf 57} (1998)
  3873--3889, [\href{http://arxiv.org/abs/physics/9711021}{{\tt
  physics/9711021}}].

\bibitem{Casas:2001sr}
J.~A. Casas and A.~Ibarra, {\it {Oscillating neutrinos and $\mu \to e,
  \gamma$}},  {\em Nucl. Phys. B} {\bf 618} (2001) 171--204,
  [\href{http://arxiv.org/abs/hep-ph/0103065}{{\tt hep-ph/0103065}}].

\bibitem{Antusch:2009gn}
S.~Antusch, S.~Blanchet, M.~Blennow, and E.~Fernandez-Martinez, {\it
  {Non-unitary Leptonic Mixing and Leptogenesis}},  {\em JHEP} {\bf 01} (2010)
  017, [\href{http://arxiv.org/abs/0910.5957}{{\tt arXiv:0910.5957}}].

\bibitem{Blennow:2011vn}
M.~Blennow and E.~Fernandez-Martinez, {\it {Parametrization of Seesaw Models
  and Light Sterile Neutrinos}},  {\em Phys. Lett. B} {\bf 704} (2011)
  223--229, [\href{http://arxiv.org/abs/1107.3992}{{\tt arXiv:1107.3992}}].

\end{thebibliography}\endgroup

\end{document}